\documentclass[twocolumn]{autart}    

\usepackage{graphicx}          

\usepackage{amsfonts, amssymb, epsfig, xspace}
\usepackage{amsmath}         

	\newtheorem{theorem}{Theorem}
	\newtheorem{assumption}{Assumption}
	\newtheorem{remark}{Remark}

	\newtheorem{lemma}{Lemma}

	\newtheorem{problem}{Problem}

\usepackage{color}

\newcommand{\todo}[1]{\textcolor{red}{{#1}}}

\newenvironment{list4}{
	\begin{list}{$\bullet$}{%
			\setlength{\itemsep}{0.05cm}
			\setlength{\labelsep}{0.2cm}
			\setlength{\labelwidth}{0.3cm}
			\setlength{\parsep}{0in} 
			\setlength{\parskip}{0in}
			\setlength{\topsep}{0in} 
			\setlength{\partopsep}{0in}
			\setlength{\leftmargin}{0.2in}}}
	{\end{list}}

\usepackage[natbibapa]{apacite}

\usepackage[font=footnotesize,skip=0pt]{caption}
\usepackage{color}
\usepackage{mathtools}

\DeclarePairedDelimiter{\diagfences}{ (  }{ ) }
\newcommand{\diag}{\operatorname{diag}\diagfences}

\usepackage{soul}
\usepackage{algorithmicx}
\usepackage{verbatim}
\begin{document}

\begin{frontmatter}
\title{Detailed Derivations of ``Multi-agent consensus with heterogeneous time-varying input and communication delays in digraphs''}  



\author[Aalto]{Wei Jiang}\ead{wei.jiang@aalto.fi}, 
\author[Beijing]{Kun Liu}\ead{kunliubit@bit.edu.cn},   
and
\author[Aalto]{Themistoklis Charalambous}\ead{themistoklis.charalambous@aalto.fi}

\address[Aalto]{Department of Electrical Engineering and Automation, School of Electrical Engineering, Aalto University, Finland}  
\address[Beijing]{School of Automation, Beijing Institute of Technology, 100081 Beijing, China} 
          

\begin{abstract}                          
This paper investigates the distributed consensus tracking control problem for general linear multi-agent systems (MASs) with external disturbances and heterogeneous  time-varying input and communication delays under a directed communication graph topology, containing a spanning tree. First, for all agents whose state matrix has no eigenvalues with positive real parts, a communication-delay-related observer, which is used to construct the controller, is designed for followers to estimate the leader's state information. Second, by means of the output regulation theory, the results are relaxed to the case that only the leader's state matrix eigenvalues have non-positive real parts and, under these relaxed conditions, the controller is redesigned. Both cases lead to a  closed-loop  error system of which the stability  is guaranteed via  a Lyapunov-Krasovskii functional with sufficient conditions in terms of input-delay-dependent linear matrix inequalities (LMIs). An extended LMI is proposed which, in conjunction with the rest of LMIs, results in a solution with a larger upper bound on delays than what would be feasible without it. It is highlighted that the integration of communication-delay-related observer and input-delay-related LMI to construct a fully distributed controller (which requires no global information) is scalable to arbitrarily large networks. The efficacy of the proposed scheme is demonstrated via illustrative numerical examples.
\end{abstract}

\end{frontmatter}

\section{Introduction}
The design of algorithms for distributed coordination in networked systems has attracted significant attention by many disciplines over the last few decades, such as control, communication, physics, biology, and computer science.
The emergence of this type of network systems, stretching from smart grids, social, robotic, and traffic networks of various sorts to embedded electronic devices, has sparked immense interest in distributed coordination problems. One such coordination problem is \emph{consensus tracking control of multi-agent systems (MASs)} in which followers are designed to track their leader; see, for example, \cite{olfati-saber_consensus_2004}.



The design of feedback control systems in MASs connected over a communication network inherits two types of delays: \emph{input} and \emph{communication} delays. Input delays (IDs) are related to the existence of communication links in the feedback control loop (sensor-to-controller delay and controller-to-actuator delay) inducing delays (due to, e.g., limited bandwidth, retransmissions, and slow processing times) while exchanging data among devices. Communication delays (CDs) are attributed to the delayed information from neighboring agents via the underlying communication network (due to retransmissions, congestion, limited bandwidth, etc). Both types of delays affect the stability of the whole system.


Many consensus controllers have been proposed to tackle \emph{homogeneous} CDs, e.g., in~\cite{zhou2014consensus}. One key advantage of addressing the problem of having homogeneous delays is the easiness to put the MAS dynamics into a compact mathematical form related to the Laplacian matrix of the communication graph. For \emph{heterogeneous} delays, however, the above advantage disappears and linear matrix inequality (LMI) conditions are often proposed, e.g., in~\cite{5109512}, to deal with the heterogeneous nature of CDs. However, these LMI conditions are not scalable to arbitrarily large networks as the dimension of the LMI increases with the number of agents or the number of the delays. Alternatively, heterogeneous fixed delays can be transformed into the Laplace domain and approaches in the frequency domain (e.g., generalized Nyquist criterion) can be  utilized to design controllers for specific dynamics of MASs, i.e., single-input-single-output~\citep{munz2010delay}, first order~\citep{ahmed2020consensus}
and general linear dynamics~\citep{jiang_LCCS}. However, time-varying delays cannot be transformed and analyzed in the frequency domain. The difficulty of standard techniques to deal with  time-varying heterogeneous {CDs (HCDs)} motivates this work.


Different from CDs, IDs have been investigated for decades for the single-agent system.
To actively compensate for IDs, predictive controllers have been proposed in the literature, e.g., the Smith predictor, the finite spectrum assignment approach, Artstein's model reduction technique and the transport partial differential equation technique; {see~\cite{fridman2014introduction_book} and references therein}.
Apart from  the Smith predictor in the frequency domain, above predictive controllers always have an integral term. As it is claimed in~\cite{833279}, \emph{the integral term discretization should be carefully executed in real applications, especially for open-loop unstable systems, because the bad discretization may make systems become unstable}. Therefore, it is beneficial to drop the integral term, not only for improving the computational efficiency, but also for not compromising the stability of the system.
\cite{Besancon2007} proposed the predictive observer approach without any integral term, which is followed in~\cite{najafi2013closed} for constant IDs and in~\cite{Lechappe2016} for time-varying IDs.
For MASs, the problem considering heterogeneous {IDs (HIDs)} emerges and is challenging as Kronecker format dynamics for MASs cannot be constructed like in the case of homogeneous ones.
To deal with this problem, there are mainly two methods: \emph{i)} One is the frequency-domain  approach for constant delays which is, e.g., utilized in~\cite{de2018trajectory} with single/double integrator dynamics. 
\emph{ii)} The other, instead of dealing with HIDs for the whole MAS using the Kronecker product method, is transforming the consensus problem into studying a single agent with its own ID, see, e.g., in~\cite{xu2017consensus} {where only the constant ID is handled and a sum term related to historical input information is needed in the discrete-time dynamics which echoes the integral term in the continuous-time dynamics}.
To the authors' best knowledge, to deal with  time-varying HIDs for general linear MASs is still an open challenge, which is the other motivation of this work.

There are also some works considering IDs and CDs simultaneously. For example, for constant IDs and CDs, see~\citep{4639466,ahmed2020consensus,jiang_LCCS}; for constant IDs but time-varying {CDs}, see~\citep{zhou2014consensus,xu2017consensus}.
To the authors' best knowledge, there is no work dealing with time-varying IDs and CDs at the same time, which is more realistic in real applications. In this paper, we aim to close this gap.

In this paper,  time-varying {HIDs and HCDs} and external disturbances are addressed for multi-agent consensus tracking under a directed graph topology. {Instead of designing an observer related to IDs and CDs together as in~\cite{jiang_LCCS}, }the idea is to decouple IDs and CDs during controller construction by designing an only\--CD\--related observer such that the ID can be dealt with inside each agent separately using the LMI technique.
The {main} contributions are as follows. 

\begin{list4}
	\item This work can deal with time-varying HIDs and time-varying HCDs at the same time.
	\item A larger upper bound of IDs is achieved by proposing a new objective-function-transformed LMI when optimizing existing LMIs.
	How to adjust this LMI to make unstable MASs become stable is also provided. 
	
	
	\item The  proposed controller is \emph{fully} distributed (no global information is needed) compared to aforementioned works in the literature dealing with delays, such as, \cite{5109512,zhou2014consensus,de2018trajectory,ahmed2020consensus} and without integral terms (which is computationally efficient and does not compromise stability). 
\end{list4}

\vspace{-0.5cm}
%
%
%
%
\section{Preliminaries and problem formulation}\label{section2}
\vspace{-0.3cm}
\textbf{Notation}. 
Throughout this paper, $\mathbb{R}^{m \times n}$ and $\mathbb{R}^{n}$ are the $m \times n$ real matrix space and the $ n $-dimensional Euclidean vector space, respectively.
$\otimes$ is the Kronecker product and $ \text{col}(\cdot) $ denotes a column vector.
$ \diag{a_1, \ldots, a_{n}} $ represents a diagonal matrix with  diagonal elements  $  a_1, \ldots, a_{n}$.
Matrices are assumed to have compatible dimensions if not explicitly stated. 
A matrix $ A \in \mathbb{R}^{n\times n} $ is called Metzler if every off-diagonal entry of $ A $ is non-negative.
$ \lambda_{\min}(A) $ and $ \lambda_{\max}(A) $ represent the minimum and maximum eigenvalues of $ A $, respectively.
The square matrix $ A \succ  0 $ ($A \succeq 0$)  means $ A $ is symmetric and  positive (semi) definite.
{$ L_{\infty}(a,b) $ is the space of essentially bounded functions $ \phi : (a,b) \rightarrow \mathbb{R}^{n}  $ with the norm $ \| \phi \|_{\infty} = \text{ess} \sup_{\theta \in (a,b)}|\phi (\theta)| $.}
For a vector $ x $, denote $ \|x\| $ as its 2-norm. 
For any integer $ a \le b $, denote $ \textbf{I}_a^b=\{a, a+1,\ldots, b\} $.
Symmetric terms in symmetric matrices are denoted by $ * $, e.g.,  $
	{\tiny \begin{bmatrix}
		A & B\\ * & C
		\end{bmatrix} = \begin{bmatrix}
		A & B\\ B^T & C
		\end{bmatrix}}
	$.\\
\textbf{Graph theory}.
In a weighted graph $ \mathcal{G} = (\mathcal{V,E,A})$, $\mathcal{V}=\{1,2, \ldots, N\}$ and $\mathcal{E} \subseteq \mathcal{V \times V}$ are respectively the nodes and edges. $\mathcal{A} = [ a_{ij} ] \in \mathbb{R} ^{N \times N}$ is the weighted adjacency matrix with $a_{ij}=1, (i,j) \in \mathcal{E}$ and $a_{ij} = 0$ otherwise. An edge $(i,j) \in \mathcal{E}$ means agent $j$ can get information from  $i$ but not necessarily conversely.
The Laplacian matrix $\mathcal{L} = \left[ l_{ij} \right] \in \mathbb{R} ^{N \times N}$ is defined as $l_{ij}= -a_{ij}, i \neq j$ and $l_{ii}= \sum_{j \neq i} a_{ij} $. 
A directed path from node $i$ to $j$ is a sequence of edges $\left(i,i_{1}\right),\left(i_{1},i_{2}\right), \ldots,\left(i_{h}, j\right)$ in the directed network with distinct nodes $i_{1}, i_2, \ldots, i_h $. 
A digraph (i.e., directed graph) contains a directed spanning tree if there is a node from which a directed path exists to each other node.\\
%
\textbf{System model}.
Consider a group of {homogeneous MASs with $ N $ followers and the leader indexed by $ 0 $} as
\begin{align}
\dot{x}_{i}(t)  = & Ax_{i}(t)+Bu_{i}(t-\tau_{u_i}(t)) + v_i(t), i \in \textbf{I}_1^N,\label{follower_dynamics}\\
\dot x_{0}(t)  = & A x_{0}(t),\label{chap_7_vl_dynamics}
\end{align}
where  $ x_{i}(t)=[x_{i1}(t), \ldots, x_{in}(t)]^T \in \mathbb{R}^n$ and $u_{i}(t)  \in \mathbb{R}^{p} $ are respectively the state and input of the $ i $-th follower and $x_{0}(t) \in \mathbb{R} ^{n}$.
The impact of an uncertain environment on each agent's dynamics is modeled by the exogenous disturbance 
$ v_i(t)\in \mathbb{R} ^{n} $ 
{which is supposed to be locally essentially bounded meaning that $  v_i (t) \in L_{\infty}[0, t), \forall t > 0 $, i.e., $ \|v_i [0,t]\|_{\infty} \le \Delta_{i} $ with $ \Delta_{i} $ is a priori given.}
$ (A,B) $ is controllable. 
Not all followers can receive the leader's state information.
$ \tau_{u_i}(t) $ is the unknown time-varying HID.
Denote the CD from agent $ j $ to agent $ i $ as $ \tau_{c_{ij}}(t) $ which can be heterogeneous and time-varying.
{$ \tau_{u_i}(t) $ and $ \tau_{c_{ij}}(t) $ satisfy the following assumptions.} 
\vspace{-0.3cm}
\begin{assumption}\label{assump_delay}
	Input delays are upper bounded ( $ 0\le \tau_{u_i}(t) \le   {\bar{\tau}_i, \bar{\tau} = \max_{i \in \textbf{I}_1^N} \bar{\tau}_i } $) and
	differentiable with their derivatives upper bounded ($ \dot{\tau}_{u_i}(t) \le  \hat{\tau}_i, \hat{\tau} = \max_{i \in \textbf{I}_1^N} \hat{\tau}_i  $).
\end{assumption}
\vspace{-0.3cm}
\begin{assumption}\label{assump_communication_delay}
	Each agent $ i $  knows the value of $ \tau_{c_{ij}}(t) $ when its neighbor agent $ j $ sends information to it.
\end{assumption}
\vspace{-0.3cm}
In several real-world applications, devices use timestamps at the transmitted packets. As a result, the receiving node $i$ is able to measure the delay $ \tau_{c_{ij}}(t) $ for a packet arriving from node $j$. Note that the assumption of  known {CDs} appears in several works in the literature (see, e.g.,~\cite{zhou2014consensus,hou2017consensus,jiang_LCCS} and references therein).
\vspace{-0.3cm}
\begin{assumption}\label{assumptiondirected}
	Graph $ \mathcal{G} $ contains a directed spanning tree in which the leader acts as the root node.
\end{assumption}  
\vspace{-0.3cm}
Then, the Laplacian matrix {$ \mathcal{L} $} of $ \mathcal{G} $ can be partitioned as $ \mathcal{L} =  
{\tiny \begin{bmatrix}
	0 & 0_{1 \times N} \\
	\mathcal{L}_{2} & \mathcal{L}_{1}  
	\end{bmatrix}} $, where $  \mathcal{L}_{2} \in \mathbb{R}^{N \times 1}$ and $ \mathcal{L}_{1} \in \mathbb{R}^{N \times N} $. 
{We denote the multi-agent set with and without the leader as $ \mathcal{N} $ and $  \mathcal{\bar N} $, respectively.}
Based on \eqref{follower_dynamics} and \eqref{chap_7_vl_dynamics}, denote the  consensus tracking error for follower $ i $ as 
$ \tilde x_{i}(t)=x_{i}(t)-x_{0}(t) $ and we have
\begin{align}
\dot{\tilde{x}}_{i}(t)  =& A\tilde{x}_{i}(t)+Bu_{i}(t-\tau_{u_i}(t))+ v_i(t), i \in \textbf{I}_1^N.  \label{chap_7_x_tilde}
\end{align}
In addition to homogeneous MASs, we also consider the heterogeneous ones as
\begin{equation}\label{mas_heterogeneous}
\begin{aligned}
\dot{x}_{i}(t)  = & A_ix_{i}(t)+B_iu_{i}(t-\tau_{u_i}(t)) + v_i(t),\\
y_{i}(t) = & C_{i}x_{i}(t), i \in \textbf{I}_1^N,\\
\dot x_{0}(t) = & A_0 x_{0}(t),
y_{0}(t) =  C_0x_{0}(t),
\end{aligned}
\end{equation}
where $ x_{i}(t)  \in \mathbb{R}^{n_i}, v_i(t)\in \mathbb{R} ^{n_i}, u_{i}  \in \mathbb{R}^{p_i}, y_{i}\left(t\right) \in \mathbb{R} ^{q}$ and $x_{0}(t) \in \mathbb{R} ^{n}$, $y_{0}(t) \in \mathbb{R} ^{q}$. $ (A_i,B_i) $ are controllable. $ C_i $ and $ C_0 $ are the output matrix. 
Here, the reason to choose $ A_0 $  for the leader instead of $ A $ is for the presentation convenience. Other variables are the same as the ones in homogeneous MASs.
We change  $ \tilde x_{i}(t)=x_{i}(t)-x_{0}(t) $ for homogeneous MASs as the output consensus error $ \tilde x_{i}(t)=y_{i}(t)-y_{0}(t) $ for heterogeneous MASs.
\vspace{-0.3cm}
\begin{problem}\label{problem}
	Considering time-varying  {HIDs and HCDs},
	for any given initial states $ x_{i}(0) \cup x_0(0) $, design a distributed controller 
	{to achieve the following objectives:
		\vspace{-0.3cm}
		\begin{itemize}
			\item[I:] the consensus tracking error $ \tilde{x}_{i}(t) $ {for homogeneous MASs \eqref{follower_dynamics} and \eqref{chap_7_vl_dynamics}} is exponentially stable if $ v_i (t) \equiv 0, i \in \textbf{I}_1^N $, and 
			stays bounded if $ v_i (t) \in L_{\infty}[0, t), \forall t > 0 $; 
			\item[II:] the output consensus error $ \tilde x_{i}(t)$ for heterogeneous MASs \eqref{mas_heterogeneous} stays bounded;
			\item[III:] MASs can endure larger delays.
	\end{itemize}}
\end{problem}
\vspace{-0.5cm}
%
%
%
%
\section{Communication-delay-related observer}\label{sec_leaderobserver}
\vspace{-0.3cm}
In this section, an observer which only involves {CDs} is designed {for homogeneous MASs} under Assumption~\ref{assump_A} in Subsection~\ref{subsec_homogeneousMAS}. In order to further relax the above constraint that 
 all agents' state matrix $ A $ has no eigenvalues with positive real parts,
the output regulation theory is deployed to get the relaxed Assumption~\ref{assump_A0} in Subsection~\ref{sec_relax_assump}, which naturally results in the consensus control for heterogeneous MASs. Note that the  CD-related observer is the first step (also the key step), to address multi-agent consensus when the time-varying IDs and CDs are considered simultaneously. 
{In the rest of this paper, for the convenience of presentation, we will omit the term $ (t) $ in $ \tau_{c_{ij}}(t) $ or $ \tau_{u_i}(t) $.
When there exists no confusion, the variable $ t $ will be omitted, e.g., $ x=x(t) $. }
\vspace{-0.3cm}
\subsection{Observer \&  controller for homogeneous MASs}
\label{subsec_homogeneousMAS}
\vspace{-0.3cm}
\begin{assumption}\label{assump_A}
		The state matrix $ A $ for MASs has no eigenvalues with positive real parts.
\end{assumption}
\vspace{-0.3cm}
In order to achieve consensus tracking, each follower should have knowledge about the leader's state information. Thus, 
design a distributed observer $ \xi_i(t) \in \mathbb{R}^{n} $ as
\vspace{-0.3cm}
\begin{align}
\dot \xi_i(t) =& A\xi_i(t) + \epsilon \sum_{j \in \mathcal{\bar N}, j \neq i} a_{ij}[ e^{A\tau_{c_{ij}}} \xi_j (t- \tau_{c_{ij}} ) - \xi_i(t)] \nonumber\\
&+ \epsilon a_{i0} [ e^{A\tau_{c_{i0}}} x_0 (t- \tau_{c_{i0}} ) - \xi_i(t)  ], i \in \textbf{I}_1^N, \label{leader_observer}
\end{align} 
where $ 0 < \epsilon \in \mathbb{R} $ is a constant{ and $ \xi_i(t)=0, t \le 0 $}.
$ \xi_j (t- \tau_{c_{ij}} ) $ denotes the communication-delayed observer information from agent $ j $ to agent $ i $, i.e., $ \xi_j (t- \tau_{c_{ij}} ) $ means agent $ j $ sends its observer information $  \xi_j (t) $ to the neighboring agent $ i $ via communication topology edge $ (i, j) $ which has communication delay $ \tau_{c_{ij}} $. 
The same holds for the leading agent $ x_0 (t- \tau_{c_{i0}} ) $.
Denote the observer estimate error is $ \tilde{\xi}_i  = \xi_{i} - x_0$.
\vspace{-0.2cm}
\begin{remark}
From the construction of observer~\eqref{leader_observer}, agent $ i $ does not need to  use a delayed value of its state, unlike, e.g.,~\cite{zhou2014consensus,hou2017consensus,jiang_LCCS} in which their results would not be feasible if an agent  does not  use a delayed value of its state. Since the receiving node $i$ is able to measure the delay $ \tau_{c_{ij}} $ for a packet arriving from node $j$ (Assumption~\ref{assump_communication_delay}),  it is able to calculate observer~\eqref{leader_observer}.
\end{remark}
\vspace{-0.3cm}
\begin{lemma}\label{theorem_tracking_input}
	Under Assumptions~\ref{assump_communication_delay}-\ref{assump_A},  $  \epsilon >0 $, the observer estimate error yields
	$ \lim_{t\rightarrow \infty} \tilde{\xi}_i(t) = 0 $ exponentially.
\end{lemma}
\vspace{-0.75cm}
\begin{pf}	
See Appendix~\ref{appendix_lemma1}.
\end{pf}
\vspace{-0.7cm}
Now, the control input is chosen to be of the form  as
\begin{equation}\label{input}
u_i(t)= K (x_{i}(t) -\xi_i(t) ), i \in \textbf{I}_1^N,
\end{equation}where {the controller gain matrix} $ K \in  \mathbb{R}^{p\times n} $ will be designed later.
Based on $ u_i= K (x_{i}-\xi_i - x_{0}+x_{0}  )
= K (\tilde x_{i} -\tilde \xi_i ) $,
integrating the above equation into~\eqref{chap_7_x_tilde} gives
\begin{equation}
\begin{aligned}
\dot{\tilde{x}}_{i}  = A\tilde{x}_{i}+BK\tilde x_{i}(t-\tau_{u_i})  + v_i- BK\tilde\xi_i(t-\tau_{u_i}).\label{x_tilde2}
\end{aligned}
\end{equation}
We regard the term $  v_i(t) - BK\tilde\xi_i(t-\tau_{u_i}) $ as the disturbance to the error dynamics~\eqref{x_tilde2}. 
As $ v_i(t) \in  {L_{\infty}[0, \infty)} $ in~\eqref{follower_dynamics}  and $ \lim_{t\rightarrow \infty} \tilde{\xi}_i(t) = 0 $ in Lemma~\ref{theorem_tracking_input}, $  (v_i(t) - BK\tilde\xi_i(t-\tau_{u_i})) \in  {L_{\infty}[0, \infty)} $.
Since \eqref{x_tilde2} is only related to agent index $ i $, thus, $ i $ will be omitted in the following. Therefore,  denote $ \zeta \coloneqq \tilde{x}_{i}, \tau(t)\coloneqq\tau_{u_i}(t),  \varpi \coloneqq v_i- BK\tilde\xi_i(t-\tau_{u_i}) $ and $ \hat{\tau}\coloneqq\hat{\tau}_i $ such that $ \dot{\tau}(t) \le  \hat{\tau}  $ from Assumption~\ref{assump_delay}. Then, the transformed error dynamics is 
\vspace{-0.3cm}
\begin{align}
\dot{\zeta}(t)=A\zeta(t) + BK\zeta(t-\tau(t)) +\varpi(t). \label{chap_7_x_tilde_Z1}
\end{align}
\vspace{-0.8cm}
\subsection{Observer \&  controller for heterogeneous MASs}\label{sec_relax_assump}
\vspace{-0.3cm}
	Results in Assumption~\ref{assump_A} 
	is restrictive for all followers and the leader. However, for relaxing this assumption, one way is that followers and the leader should have different state matrix $ A $, i.e., the system changes to heterogeneous MAS {as in~\eqref{mas_heterogeneous}}. 
	 Then, the following assumption based on output regulation theory in~\cite{huang2004nonlinear} is needed.
	 \vspace{-0.3cm}
\begin{assumption}\label{assumption_regulation}
	There exist solutions $ (X_{i}, U_{i}) $ for each follower $ i$ to
		the following linear matrix equations
		\begin{align}\label{output_regulation1}
		X_{i}A_0  =  A_{i}X_{i}+B_{i}U_{i}, 
		C_0  =  C_{i}X_{i}, i \in \textbf{I}_1^N. 
		\end{align}
	\end{assumption}
\vspace{-0.3cm}
	\begin{assumption}\label{assump_A0}
		{Eigenvalues of the} leader's state matrix $ A_0 $ have {either the following properties: i)  negative real parts; ii) zero real part but are simple, i.e., eigenvalues on the imaginary axis are all distinct from one another.}
	\end{assumption}
\vspace{-0.3cm}
	Based on Assumption~\ref{assumption_regulation}, Assumption~\ref{assump_A}  can be relaxed to Assumption~\ref{assump_A0}{ in which the leader dynamics is asymptotically stable or  marginally stable [Theorem 8.1, \cite{hespanha2018linear}]}. The motivation behind this is that several real-world scenarios may involve follower state dynamics $ A_i $ that are open-loop unstable (e.g., fight aircrafts).
	The distributed observer in \eqref{leader_observer} is thus changed to 
	\vspace{-0.3cm}
	\begin{align}
	\dot \xi_i&(t) = A_0\xi_i(t) + \epsilon \sum_{j \in \mathcal{\bar N}, j \neq i} a_{ij}[ e^{A_0\tau_{c_{ij}}} \xi_j (t- \tau_{c_{ij}} )\label{leader_observer_heterogeneous}\\
	& - \xi_i(t)] + \epsilon a_{i0} [ e^{A_0\tau_{c_{i0}}} x_0 (t- \tau_{c_{i0}} ) - \xi_i(t)  ],i \in \textbf{I}_1^N  \nonumber
	\end{align} 
	with $ \xi_i(t) \in \mathbb{R}^{n} $.
	The difference is the replacement of $ A $ in~\eqref{leader_observer} to $ A_0 $ in~\eqref{leader_observer_heterogeneous}. Therefore, $ \lim_{t\rightarrow \infty} \tilde{\xi}_i(t) = 0 $ is still valid under Assumptions~\ref{assump_communication_delay}, \ref{assumptiondirected} and \ref{assump_A0} with the necessary condition for the positive parameter $ \epsilon $ as 
	$ \mathrm{Re} ({\lambda} (I_{N}\otimes A_0  -\mathcal{L}_1\otimes (\epsilon I_n) ) )   <0 $.
	We redesign the control input as 
	\begin{equation}\label{input_heterogeneous}
	u_i(t)= K^1_i x_{i}(t) {-}  K^2_i \xi_i(t), i \in \textbf{I}_1^N,
	\end{equation}
	where the controller gain matrices $ K^2_i = U_i - K^1_iX_i $ and $ K^1_i \in \mathbb{R}^{p_i \times n_i} $ will be designed later.
 Denote $ \bar x_i = x_i - X_ix_0 $. Based on~\eqref{output_regulation1}, we have
	\begin{equation}\label{output_consensus_error}
	\tilde x_{i}= y_i-y_0 = C_i(\bar x_i + X_ix_0 ) - C_0x_0 = C_i\bar x_i,
	\end{equation}
	which means the output consensus error $ \tilde x_{i} $ is dependent on the term $ \bar x_{i} $. 
	Based on Eqs.~\eqref{mas_heterogeneous}, \eqref{output_regulation1} and \eqref{input_heterogeneous}, the derivative of  $ \bar x_{i} $ is calculated as
	\begin{align}
	\dot{\bar x}_i 
	=& A_i\bar x_i + B_i K^1_i \bar x_i(t-\tau_{u_i})+ v_i- B_i K^2_i\tilde\xi_i(t-\tau_{u_i})\nonumber\\
	& - B_iU_i( x_0(t)  -x_0(t-\tau_{u_i})). \label{calculation_bar_xi}
	\end{align}
	One can see \eqref{calculation_bar_xi} has a similar math format as \eqref{x_tilde2}. Similarly, denote  $ \zeta \coloneqq \bar{x}_{i}, \tau(t)\coloneqq\tau_{u_i}(t),  \varpi \coloneqq  v_i- B_i K^2_i\tilde\xi_i(t-\tau_{u_i}) - B_iU_i( x_0(t ) -x_0(t-\tau_{u_i}))$, $ \hat{\tau}\coloneqq\hat{\tau}_i $  and $ A \coloneqq A_i, B \coloneqq B_i, K \coloneqq K^1_i $. Then, \eqref{calculation_bar_xi} changes to \eqref{chap_7_x_tilde_Z1}. 
	{We should verify whether $ \varpi \in L_{\infty}[0, \infty) $. Denote $ \varrho_i(t) = x_0(t) -x_0(t-\tau_{u_i}) $; then, from~\eqref{mas_heterogeneous} it is easy to get $ \dot{\varrho}_i = A_0 \varrho_i + \dot{\tau}_{u_i}A_0x_0(t-\tau_{u_i}) $. 
	Denote another augmented variable $ \bar  \varrho_i(t) = [\varrho^T_i(t), x^T_0(t-\tau_{u_i})]^T $ and we have $ \dot{\bar  \varrho}_i = {\tiny \begin{bmatrix}
			A_0& \dot{\tau}_{u_i}A_0\\0&A_0
			\end{bmatrix}} \bar  \varrho_i = ({\tiny \begin{bmatrix}
			1& \dot{\tau}_{u_i}\\0&1
			\end{bmatrix}} \otimes A_0)\bar  \varrho_i $. Based on the fact that suppose the eigenvalues of $ S\in \mathbb{R}^{n\times n} $ and $ T\in \mathbb{R}^{m\times m} $ are $ \lambda_{1},\ldots,\lambda_{n} $ and $ \mu_{1},\ldots, \mu_{m} $, respectively, then the eigenvalues of $ S\otimes T $ are $ \lambda_{i}\mu_{j}, i=1,\ldots,n, j=1,\ldots, m $, one can verify that the stability of $ \bar  \varrho_i(t) $ is determined by the eigenvalues of $ A_0 $.}
	Thus, based on Assumption~\ref{assump_A0}, {$ \bar  \varrho_i(t) $ dynamics is asymptotically or marginally stable, i.e., $ \varrho_i(t) $ is bounded.}  Therefore, $ \varpi(t) \in   {L_{\infty}[0, \infty)} $ is still valid here. Finally, $ K^1_i $ will be designed as $ K $ in~\eqref{chap_7_x_tilde_Z1}  such that $ \bar x_i(t)  $ is bounded.
	Based on~\eqref{output_consensus_error},  $ \tilde{x} _i(t)$ for heterogeneous MASs~\eqref{mas_heterogeneous} will be accordingly also bounded.

%
%
%
%
\section{Stability analysis}\label{sec_mismatched}
\vspace{-0.3cm}
From the previous section, one can see that by taking advantage of designing the only CD-related observer $ \xi_i(t) $ for either homogeneous or heterogeneous MASs, 
	both the original cooperative consensus tracking problem is transformed into the  {input-to-state stability} problem of ``single agent system''~\eqref{chap_7_x_tilde_Z1} involving only the time-varying ID.
	The  LKF with descriptor method~\citep{fridman2014introduction_book} will be adopted to design  $ K $.
Inspired by the work of \cite{fridman2014introduction_book}, design one type of LKF as
\vspace{-0.7cm}
{\small \begin{align}
	&V
	=  \zeta^{T}P \zeta + \int_{t-\bar{\tau}}^{t} e^{2\delta(s-t)} \zeta^{T}(s)S\zeta(s)\text{d}s +  \int_{t-\tau(t)}^{t} e^{2\delta(s-t)} \nonumber \\& \times \zeta^{T}(s)Q\zeta(s)\text{d}s + \bar{\tau} \int_{-\bar{\tau}}^{0} \int_{t+\theta}^{t} e^{2\delta(s-t)} \dot{\zeta}^{T}(s)R\dot{\zeta}(s)\text{d}s\text{d}\theta,\nonumber
	\end{align}}\noindent
where $ 0 \le \tau(t) \coloneqq \tau_{u_i} \le \bar{\tau} $ from Assumption~\ref{assump_delay}, $ \delta >0$  is a constant and  matrices $ P \succ 0, Q {\succeq} 0, R \succ 0, S \succ 0 $.
So $ V $ is positive. 
Denote a scaler $ \gamma > 0  $ and
\begin{equation}\label{Lyapunov_W}
	W = \dot{V} + 2\delta V - \varpi^T\gamma\varpi.
	\end{equation}
	The calculation of $ W $ is presented in Appendix~\ref{appendixA}.
	Following the Proposition 1 in~\cite{fridman2009control}, if 
	there exist $ \delta >0, \gamma > 0 $ and  matrices $ \{P,S,R\} \succ 0, Q \succeq 0 $ such that along the trajectories of \eqref{chap_7_x_tilde_Z1}, the LKF satisfies the condition $ W<0 $ (i.e., the matrix inequality $ \Phi_1 $ in~\eqref{lmi_phi4} satisfies $ \Phi_1 \prec  0 $ and \eqref{S12} is feasible),
	then, the solution of error dynamics \eqref{chap_7_x_tilde_Z1} 
	satisfies
\vspace{-0.3cm}
\begin{align}
\zeta^{T}(t)P \zeta(t) \le& e^{-2\delta t} \zeta^{T}(0)P \zeta(0)+ (1-e^{-2\delta t}) \nonumber \\
&\times \frac{\gamma}{2\delta}\|\varpi[0,t]\|_{\infty}^{2}, t >0. \label{chap_7_error1}
\end{align}
\vspace{-0.7cm}
\begin{remark}\label{decayRate_delay}
	The reason for adopting the descriptor method ($ P_2, P_3 $ in~\eqref{W_calculation}) is that the controller parameter $ K $ can be designed conveniently and that some comparison simulations in Section 6.1.3 of~\cite{jiang2018fully} shows that the closed-loop system can endure larger delays with the descriptor method used. It also shows that there is a trade-off between the exponential convergence rate $ \delta $ and upper bound $ \bar{\tau}$: the larger the rate $ \delta $, the smaller the upper bound $ \bar{\tau} $. {$ Q = 0 $ means the system can endure the fast-varying delay (i.e., $ \dot{\tau}_{u_i}(t) \ge 1 $) as the derivative upper bound $ \hat \tau $ will disappear in $ \Phi_1 (4,4) $~\citep{fridman2014introduction_book}. }
\end{remark}
\begin{lemma}\label{chap_7_theorem}
		Under Assumptions~\ref{assump_delay}-\ref{assump_A} (or Assumptions~\ref{assump_delay}-\ref{assumptiondirected}, \ref{assumption_regulation}-\ref{assump_A0}),
	given $ \bar{\tau} \ge 0, \hat{\tau} \in [0,1),  \delta >0 $, $ \gamma  >0 $ and $ \varepsilon \in \mathbb{R} $,
	if there exist $ n \times n $ matrices $\{ \bar P, \bar Q,\bar R,\bar S\} \succ 0 ${ ($\bar Q=0 $ for $ \hat{\tau} \ge 1 $)}, $ \{\bar S_{12}, M \} \in  \mathbb{R}^{n \times n}  $, $ Y \in  \mathbb{R}^{p \times n  }  $ such that the following LMIs are feasible: 
	\vspace{-0.7cm}
	{\small
	\begin{align}
	\Phi_2 =&\begin{bmatrix}
	\Phi_2 (1,1)&\Phi_2 (1,2) & e^{-2\delta \bar{\tau}} \bar S_{12}  &\Phi_2 (1,4) &  I_n\\
	*&\Phi_2 (2,2)&0& \varepsilon BY&\varepsilon I_n\\
	*&*&\Phi_2 (3,3) & \Phi_2 (3,4) & 0\\
	*&*&*&\Phi_2 (4,4) &0\\
	*&*&*&*&-\gamma I_n
	\end{bmatrix} \prec 0 \label{lmi_phi5},\\
	\Phi_2^{'} \succeq& 0, \label{chap_7_lmi_S12}
	\end{align}
	\begin{equation*}
	\begin{aligned}
	\Phi_2 (1,1) =&2\delta \bar P+ \bar S+\bar Q-e^{-2\delta \bar{\tau}}\bar  R+AM+M^TA^T,\\
	\Phi_2 (1,2) =& \bar P-M+\varepsilon M^TA^T,\Phi_2 (3,4)=e^{-2\delta \bar{\tau}} (\bar R-\bar S_{12}^T),\\
	\Phi_2 (1,4) =&  BY+ e^{-2\delta \bar{\tau}} (\bar R-\bar S_{12}),\\
	\Phi_2 (2,2) =& \bar{\tau}^{2}\bar R-\varepsilon M^T - \varepsilon M,
	\Phi_2 (3,3) = -e^{-2\delta \bar{\tau}}(\bar S+\bar R),\\
	\Phi_2 (4,4) =& -(1-\hat{\tau})e^{-2\delta \bar{\tau}}\bar Q+ e^{-2\delta \bar{\tau}} (-2\bar R+\bar S_{12}+\bar S_{12}^T),
	\end{aligned}
	\end{equation*}$ \Phi_2^{'}=\begin{bmatrix}
	\bar R & \bar S_{12}\\ * & \bar R
	\end{bmatrix} $,
}\noindent 
	 then, {the objective I (II) of}
	Problem~\ref{problem} 
	is solved  by the  distributed controller consisting of input \eqref{input} and observer~\eqref{leader_observer} ( input \eqref{input_heterogeneous} and observer~\eqref{leader_observer_heterogeneous}).
	The controller gain matrix is thus designed as $  K=YM^{-1}$ ( $ K_i^1=YM^{-1}, K^2_i = U_i - K^1_iX_i,  (X_{i}, U_{i}) $ is the solution to the output regulation equation~\eqref{output_regulation1}).
\end{lemma}
\begin{pf}
	For objective I, recalling $ \zeta \coloneqq \tilde{x}_{i}, \varpi \coloneqq v_i- BK\tilde\xi_i(t-\tau_{u_i}) $ and $ \lim_{t\rightarrow \infty} \tilde{\xi}_i(t) = 0 $ exponentially, based on~\eqref{chap_7_error1}, if matrix inequalities $ \Phi_1 {<0} $~\eqref{lmi_phi4} and \eqref{S12}  are feasible, then $ \lim_{t\rightarrow \infty} \tilde{x}_{i}(t) = 0 $ exponentially if $ v_i(t) \equiv 0 $; otherwise, 
\vspace{-0.5cm}
	{\small \begin{equation}\label{chap_7_error2}
		\|\tilde{x}_{i}(t) \|^2 \le \frac{\gamma}{2\delta \lambda_{\min}(P)} \Delta_{i}^{2}, t \rightarrow \infty.
		\end{equation}
	}\noindent
	For objective II, similarly, 
	as $  t \rightarrow \infty$,
	\vspace{-0.7cm}
	{\small \begin{equation}\label{chap_7_error3}
		\|\tilde{x}_{i}(t) \|^2 \le \frac{\gamma \|C_i\|^2}{2\delta \lambda_{\min}(P)} (\Delta_{i}^{2} + \|B_iU_i( x_0  -x_0(t-\tau_{u_i}))\|^2).
		\end{equation}}\noindent
The problem left is to calculate the controller gain matrix $ K $.
	Recall that the decay rate $ \delta $, which is related to $ \bar{\tau} $ in Remark~\ref{decayRate_delay}, is a system convergence requirement and should be set in advance, meaning $ \delta $ is known.
	As $ K $ in \eqref{chap_7_x_tilde_Z1} is unknown, $ \Phi_1 $ contains two nonlinear terms: $ P_2^T BK $ and $ P_3^T BK $, since $ P_2, P_3 $ are also unknown.  From descriptor method, by setting $ P_3=\varepsilon P_2 $,
	nonlinear matrix inequality $ \Phi_1 $  has then one nonlinear term  $ P_2^T BK $. 
	Denote $ M \coloneqq P_2^{-1},  \bar P \coloneqq M^TP M, \bar Q \coloneqq M^TQM, \bar R \coloneqq M^TRM, \bar S \coloneqq M^TSM, \bar S_{12} \coloneqq M^TS_{12}M  $.
	Note that from the construction of $ \Phi_2 (2,2) $, the feasibility of $ \Phi_2 $ guarantees that $  M  $ or $ P_2 $ is positive definite. {Then, inspired from~\cite{liu2020networked}, multiplying $ \Phi_1 $ in~\eqref{lmi_phi4} by $ \diag{M^T, M^T, M^T, M^T, I_n} $ and $ \diag{M, M, M, M, I_n} $ from the left and right side, respectively, and denoting $ Y = KM $, 
		$ \Phi_1 $ in~\eqref{lmi_phi4} is linearized as LMI $ \Phi_2 $ in~\eqref{lmi_phi5}. 
		Similarly, $ \Phi_1^{'} $ in \eqref{S12} changes to $ \Phi_2^{'} $ \eqref{chap_7_lmi_S12}.}
	After $ M $ and $ Y $ are calculated through LMIs \eqref{lmi_phi5} and \eqref{chap_7_lmi_S12}, one has $  K=YM^{-1} $.	\\
	{For the objective II in heterogeneous MASs, we give each matrix with a subscript $ i $, e.g., replacing $ A, P $ as $ A_i, P_i, i \in \textbf{I}_1^N $ and solve all LMIs together.}
	 $ \blacksquare $
\end{pf}
\vspace{-0.4cm}
The application of Lemma~\ref{chap_7_theorem} is summarized as follows.
\vspace{-0.4cm}
\begin{algorithm}[h]
	\caption{Controller design for upper bound $  \bar{\tau} $
	 }
	\label{alg:MaxiMinNSGA} 
	\begin{algorithmic}[1]
		\State \textbf{Input:} 
		 Delay derivative bound $  \hat{\tau} $ and decay rate $ \delta $ based on system specifications ($ \delta $ determines how fast (exponentially) the MAS converges).
		\State \textbf{Initialization:} Set parameters $ \epsilon $ in~\eqref{leader_observer} and $ \varepsilon$.
		\State Define $ \bar P,\bar Q,\bar R,\bar S, \bar S_{12}  $ and $\gamma, M, Y $ as variables.
		\State Solve LMIs \eqref{lmi_phi5} and \eqref{chap_7_lmi_S12}  to get  $ M $ and $ Y $. 
		\State \textbf{Output:} Compute input parameter  $  K=YM^{-1} $.
	\end{algorithmic}
\end{algorithm}
\vspace{-0.7cm}
%
%
%
%
\section{Delay size analysis}\label{sec_delayanalysis}
\vspace{-0.3cm}
{This section is for the objective III of Problem~\ref{problem},} i.e.,  analyzing how to get an improved upper bound $ \bar{\tau} $. As stated in Remark~\ref{decayRate_delay}, the larger the rate $ \delta $, the smaller the upper bound $ \bar{\tau} $. Herein, $ \delta $ is predefined and fixed.\\
Different delay-dependent LMI conditions (e.g., LMIs \eqref{lmi_phi5}, \eqref{chap_7_lmi_S12}) are successfully proposed to prove the stability of closed-loop system for different control scenarios; see, e.g., ~\cite{Besancon2007,najafi2013closed,fridman2014introduction_book}.
	However, the aforementioned methods do not perform a delay size analysis and the derived conditions are usually restricted to relatively small delays.

The main idea is as follows. If  the upper bound $ \bar{\tau} $ is so large that LMI \eqref{lmi_phi5} is not feasible, it means that the tracking error will diverge, i.e., $ \|\tilde{x}_i(t)\| $ in~\eqref{chap_7_error2}/\eqref{chap_7_error3} will diverge. It also means in this situation, the value of $ \gamma/(2\delta \lambda_{\min}(P)) $ will be very large{ (check the analysis of Fig.~\ref{algorithm1_perfect} (d)-(f))}.
So the objective of minimizing $ \gamma/(2\delta \lambda_{\min}(P)) $ {and keeping $ \|\tilde{x}_i(t)\| $ in~\eqref{chap_7_error2}/\eqref{chap_7_error3} stable simultaneously} is preferred. However, the eigenvalue function of unknown LMI variables (e.g., $  P $) is not available, thus some transformation is needed.
In addition to the predefined $ \delta $, only $ \lambda_{\min}(P) $ needs attention. On the contrary to the common setting of $ P\succ 0 $, design $ P \succeq  I_n/\mu,  \mu >0 $
such that $ \lambda_{\min}(P) \ge 1/\mu $. Then, $ \gamma/(2\delta \lambda_{\min}(P)) \le (2\gamma\mu)/(4\delta ) $.
	In this way, based on the law $ 2\gamma\mu \le \gamma^2 + \mu ^2 $, the minimizing objective can change to an optimization constraint, i.e., by asking 
	\begin{equation*}
	\frac{\gamma}{2\delta \lambda_{\min}(P)}
	\le \frac{2\gamma\mu}{4\delta}
	\le \frac{\gamma^2 + \mu ^2}{4\delta}
	  \le \chi, \chi>0
	\end{equation*}
	 with $ \chi $ predefined, the constraint becomes
	 \vspace{-0.3cm}
	\begin{equation}\label{constriant}
	\gamma^2 + \mu ^2 - 4\delta  \chi \le 0
	\end{equation}
	with adding $ \mu $ as an LMI variable, which can be transformed into an LMI $ \Phi_3 = $
	\vspace{-0.3cm}
	 {\tiny $ \begin{bmatrix}
		4\delta \chi &  \gamma  & \mu\\ \gamma & 1 & 0\\ \mu & 0 &1
		\end{bmatrix} $} satisfying
\begin{equation}\label{lmi_add}
\Phi_3 \succeq 0.
\end{equation} One can see \eqref{constriant} can be transformed as $ \| [\gamma, \mu]^T \|_{2}\le 2\sqrt{\delta  \chi } $, which is a second-order cone constraint. As semi-definite programming (SDP) contains second-order cone programming (SOCP), a second-order cone constraint~\eqref{constriant} can be written as an LMI~\eqref{lmi_add}.\\
From $ M =P_2^{-1},  \bar P = M^T P M $ and $ P \succeq \frac{1}{\mu} I_n $ we have $ \bar P -   M^T (\frac{1}{\mu} I_n)  M \succeq 0 $. In addition to $ \mu I_n \succ 0 $, by Schur complement lemma, we get $ \Phi_4 = {\tiny \begin{bmatrix}
	\bar P & M^T\\ * & \mu I_n
	\end{bmatrix}} \succeq 0 $. Then, condition $ P \succeq  I_n/\mu,  \mu >0 $ changes to
\begin{equation}\label{barP}
\Phi_4 \succeq 0,  \mu >0.
\end{equation}
\vspace{-0.7cm}
\begin{theorem}
	{Based on Lemma~\ref{chap_7_theorem}, additionally $ \chi >0 $, by tuning $ \chi $} if LMIs  \eqref{lmi_phi5}, \eqref{chap_7_lmi_S12}, \eqref{lmi_add} and \eqref{barP} are feasible, then, for Problem~\ref{problem}, the objectives I, II are solved and the MAS can endure larger delay upper bound $ \bar \tau $ compared to Lemma~\ref{chap_7_theorem}, i.e., objective III is solved.
\end{theorem}
\begin{pf}
{The proof follows the same lines of the one of Lemma~\ref{chap_7_theorem}. The difference is that now, instead of $ P \succ 0 $, we add LMIs \eqref{lmi_add} and \eqref{barP} here. 
	The intuition explanation is that by adding an objective-function-transformed constraint (i.e., \eqref{constriant}) on the right-hand side of consensus tracking error $ \|\tilde{x}_i(t)\| $~\eqref{chap_7_error2}/\eqref{chap_7_error3}, and by tuning the value of $ \chi $ which provides a freedom to control the bound of $  \gamma/(2\delta \lambda_{\min}(P)) $ compared to Algorithms~\ref{alg:MaxiMinNSGA} which cannot control that bound,
	the value of $ \|\tilde{x}_i(t)\| $ will be more difficult to diverge or become large, i.e., the  system could endure larger delay size.} $ \blacksquare $
\end{pf}
\vspace{-0.3cm}
\begin{algorithm}[t]
	\caption{Controller design for larger upper bound $  \bar{\tau} $.}\label{alg:MaxiMinNSGA2} 
	\begin{algorithmic}[1]
		\State Same steps 1-2 in Algorithm~\ref{alg:MaxiMinNSGA}{ and $ \chi >0 $ in~\eqref{constriant}}.
		\State Define $ \bar  P,\bar Q,\bar R,\bar S,\bar  S_{12}  $ and $\gamma,  M, Y,\mu $ as variables.
		\State Solve LMIs  \eqref{lmi_phi5}, \eqref{chap_7_lmi_S12}, \eqref{lmi_add} and \eqref{barP}  to get $ M $ and $ Y ${ by tuning the value of $ \chi $}.
		\State \textbf{Output:} Compute input parameter $  K=YM^{-1} $.
	\end{algorithmic}
\end{algorithm}
\vspace{-0.3cm}
\begin{remark}
	Algorithms~\ref{alg:MaxiMinNSGA}-\ref{alg:MaxiMinNSGA2} can  be applied to single agent system.
	Unlike~\cite{5109512}, the dimension of proposed LMIs do not related to agent number $ N $ or delay number $ n $, thus will not increase when $ N $ or $ n $ increases. {It means Algorithms~\ref{alg:MaxiMinNSGA}-\ref{alg:MaxiMinNSGA2} are also scalable to a large number of agents.}
	{Different from Algorithm~\ref{alg:MaxiMinNSGA} that for a fixed delay upper bound $ \bar \tau $, $ K $ will be calculated accordingly and be fixed, i.e., the controller is fixed, Algorithm~\ref{alg:MaxiMinNSGA2} can calculate different $ K $ for a fixed $ \bar \tau $ by tuning $ \chi $ to control the upper bound of $ \gamma/(2\delta \lambda_{\min}(P)) $, i.e., tuning $ \chi $ offers a freedom for Algorithm~\ref{alg:MaxiMinNSGA2} in terms of a fixed $ \bar \tau $. The $ \chi $ tuning mechanism is described for different $ \bar \tau $ in Fig.~\ref{algorithm1_perfect} (d)-(f) and for a fixed $ \bar \tau $ in Fig.~\ref{algorithm2_gammaComparison}.}
\end{remark}

%
%
%
%

\section{Simulations}
\label{sec_simulations}
\vspace{-0.4cm}


\begin{figure}[!t]
	\centering
	\includegraphics[width=0.8\hsize]{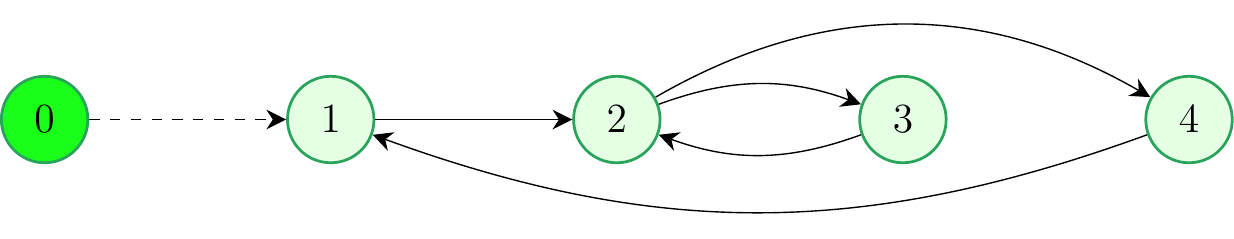}
	\caption{The digraph $ \mathcal{G} $ satisfying Assumption~\ref{assumptiondirected}.} 
	\label{chap_7_graph}
\end{figure}



Heterogeneous MASs are considered here with the graph $ \mathcal{G} $ shown in Fig. \ref{chap_7_graph}. Set the dynamics of agents $ 1 $ and $ 4 $ as the platooning dynamics in simulation of~\cite{jiang_LCCS}, i.e., $
A_1=A_4 =  
\begin{bmatrix}
0& 1& 0\\ 0& 0& 1\\ 0& 0& -2
\end{bmatrix},
B_1=B_4 =\begin{bmatrix}
0\\0\\ 2
\end{bmatrix}.
$
 Set agents $ 2 $ and $ 3 $ respectively as the linearized mobile vehicle and the Caltech wireless tested vehicle in simulation of~\cite{jiang_tac}, i.e.,
 \begin{align*}
  A_{2} & = 
 \begin{bmatrix}
 0_{2\times2} & I_{2}\\
 0_{2\times2} & 0_{2\times2}
 \end{bmatrix}, \,
 B_{2} =\begin{bmatrix}
 0_{2\times2}\\
 I_{2}
 \end{bmatrix}, \\
  A_{3} &=  
 \begin{bmatrix}
 0_{3 \times 3} & I_{3}\\
 \begin{matrix}
 0 & 0 & -0.2003 \\
 0 & 0 & 0.2003 \\
 0 & 0 & 0
 \end{matrix} & 
 \begin{matrix}
 0.2003 & 0 & 0\\
 0 & 0.2003 & 0\\
 0 & 0 & -1.6129
 \end{matrix}
 \end{bmatrix},\\
  B_{3} &=\begin{bmatrix}
 0 & 0 & 0 & 0.9441 & 0.9441 & -28.7097 \\
 0 & 0 & 0 & 0.9441 & 0.9441 & 28.7097 
 \end{bmatrix}^{T}.
 \end{align*}
 
  Set the output matrix as $ C_1=C_4=[I_2, 0_{2\times1}], C_2=[I_2, 0_{2\times2}], C_3=[I_2, 0_{2\times4}]
$.
Denote $ \bar v = [\sin (10t), \cos (10t), \\\sin (20t)]^T $ and 
set the disturbances as $ v_i(t)=0, t\in [0,200),i \in \textbf{I}_1^4; v_{1}(t) = 13 \bar v, v_{2}(t) = [2\bar v; 0], v_{3}(t) = [3\bar v; 0;0;0], v_{4}(t) = \bar v, t\in [200,400] $.

Set $ \hat{\tau}=0.8, \delta = 0.1, \varepsilon = 0.3, \epsilon = 0.3 $. 
Initial conditions are 
randomly set 
and $ u_{i}(t)=0, t\in [-\bar{\tau}, 0], i \in \textbf{I}_1^4$. 
Set the  communication delays as $ \tau_{c_{10}} = 6+ \sin(0.5t), \tau_{c_{21}} = 6+ 2\sin(0.4t), \tau_{c_{23}} = \tau_{c_{32}} = 6+ 3\sin(0.5t), \tau_{c_{42}} = 6+ 4\sin(2t), \tau_{c_{14}} = 6+ 5\sin(0.1t)$.
Set the input delays as 
$ \tau_{u_{1}}=\tau_u + 0.1\cos(0.5t), \tau_{u_{2}}=\tau_u + 0.2\sin(0.5t), \tau_{u_{3}}=\tau_u + 0.3\cos(0.1t), \tau_{u_{4}}=\tau_u + 0.4\sin(0.5t) $ with $ \tau_u \ge 0.4 $ guaranteeing $ \tau_{u_{i}} 
\ge 0, i \in \textbf{I}_1^4 $. It also means $ \bar{\tau} = \tau_u + 0.4 $ which satisfies $ \bar{\tau} \ge \tau_{u_i}  $. In the following, by changing the value of $ \tau_u $, the upper bound $ \bar{\tau} $ can be found and comparison simulations can be provided. 

\begin{figure}[h]
	\centering
	\includegraphics[width=\hsize]{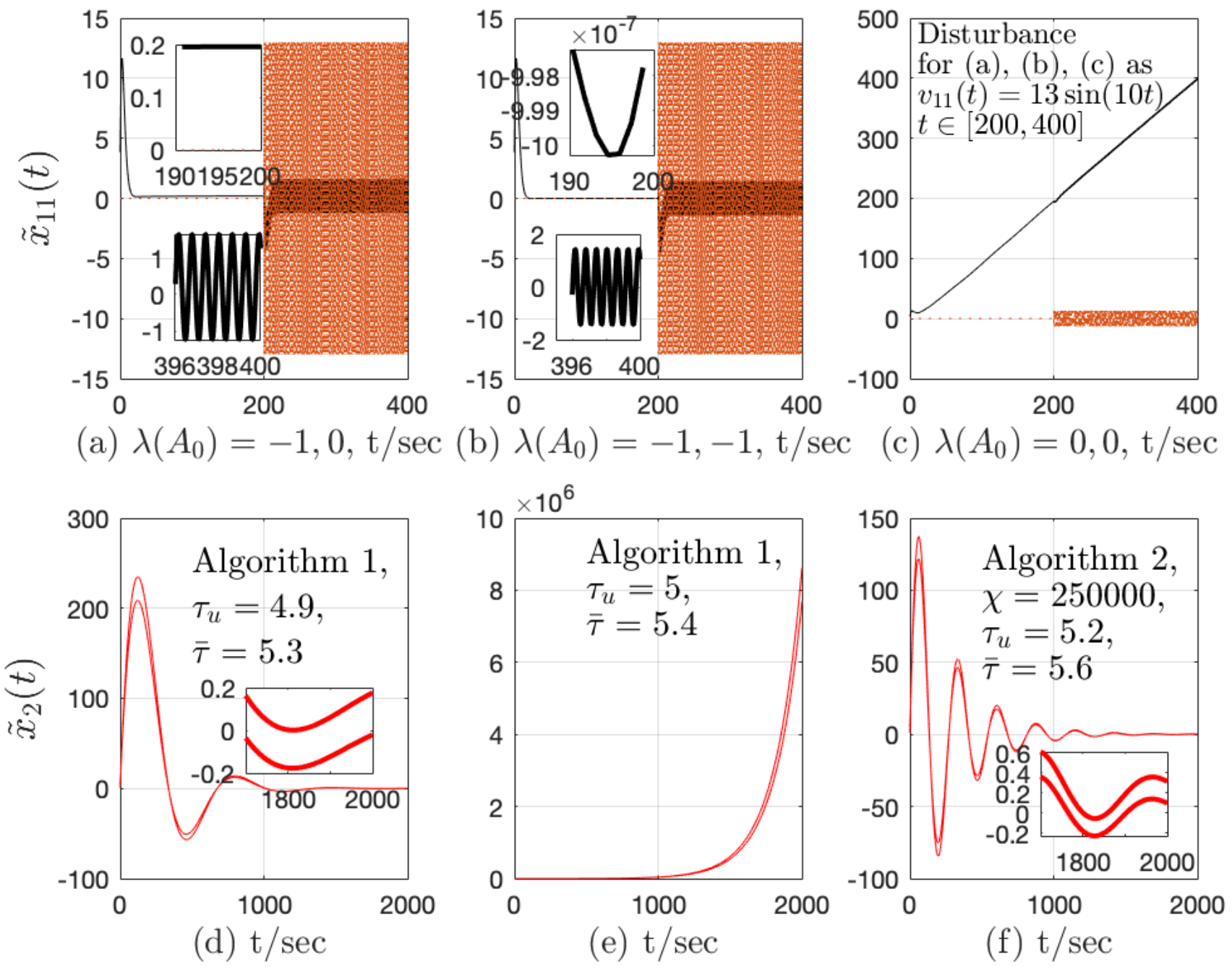}
	\caption{(a), (b), (c) are comparison of output consensus tracking error of agent $ 1 $  for different leader dynamics $ A_0 $ using Algorithm~\ref{alg:MaxiMinNSGA} with $ \tau_u=0.5, \bar{\tau}=0.9 $. (d), (e), (f) are Algorithm~\ref{alg:MaxiMinNSGA}-\ref{alg:MaxiMinNSGA2}  comparison related to the input delay bound $ \bar \tau $.
	}
	\label{algorithm1_perfect}
\end{figure}

To verify Assumption~\ref{assump_A0}, we set $ A_0 $ as marginally stable, asymptotically stable and unstable as $ A_0 = {\tiny \begin{bmatrix}
	0& 1\\ 0& -1
	\end{bmatrix}, \begin{bmatrix}
	-1& 1\\ 0& -1
	\end{bmatrix}, \begin{bmatrix}
	0& 1\\ 0& 0
	\end{bmatrix}  } $, respectively. The eigenvalues of $ A_0 $ are correspondingly shown in Fig.~\ref{algorithm1_perfect} (a), (b), (c). The solutions $ (X_{i}, U_{i}) $ to the output regulation equation~\eqref{output_regulation1} are thus obtained by using [Lemma 4, \cite{cai2017adaptive}]. As a result, for $ A_0 = {\tiny \begin{bmatrix}
	0& 1\\ 0& -1
	\end{bmatrix}}$, we get
\begin{align}
X_1 =& X_4 =\begin{bmatrix}
1  & 0\\
0 &   1\\
0  & -1
\end{bmatrix}, U_1 =U_4 =\begin{bmatrix}
0 &-0.5
\end{bmatrix},\nonumber\\
X_2 =& \begin{bmatrix}
1  & 0\\
0   &  1\\
0   &  1\\
0   & -1
\end{bmatrix}, U_2 =\begin{bmatrix}
0   & -1\\
0   &  1
\end{bmatrix},\\
X_3 =& \begin{bmatrix}
1 &   0\\
0 &   1\\
0 &   3.9925\\
-0 &   1\\
0  & -1\\
-0 &  -3.9925
\end{bmatrix}, U_3=\begin{bmatrix}
0&    0.0426\\
0  & -0.0426
\end{bmatrix}.\nonumber
\end{align}
For $ A_0 = {\tiny \begin{bmatrix}
	-1& 1\\ 0& -1
	\end{bmatrix}}$, we get
\begin{align*}
X_1 = X_4 =\begin{bmatrix}
0.7273  & -0.0909\\
-0.3637 &   0.9090\\
0.3631  & -1.2731
\end{bmatrix}, 
\end{align*}
\begin{align}
U_1 =&U_4 =\begin{bmatrix}
0.1822  & -0.4540
\end{bmatrix},\nonumber\\
X_2 =& \begin{bmatrix}
1  &  0\\
0  &  1\\
-1 &   1\\
0  & -1
\end{bmatrix}, U_2 =\begin{bmatrix}
 1  & -2\\
0  &  1
\end{bmatrix},\\
X_3 =& \begin{bmatrix}
1  & -0\\
0  &  1\\
-1.9963  &  6.4888\\
-1 &   1\\
0  & -1\\
1.9963  & -8.4850
\end{bmatrix}, U_3=\begin{bmatrix}
0.1904  & -0.2091\\
0.2331  & -0.3205
\end{bmatrix}.\nonumber
\end{align}
For $ A_0 = {\tiny \begin{bmatrix}
	0& 1\\ 0& 0
	\end{bmatrix}}$, we get
\begin{align}
X_1 =& X_4 =\begin{bmatrix}
1   &      0\\
0  &  1\\
0  &  0
\end{bmatrix}, U_1 =U_4 =\begin{bmatrix}
0 &0
\end{bmatrix},\nonumber\\
X_2 =& \begin{bmatrix}
1     &    0\\
0.0001  &  1\\
0.0001  &  1\\
0.0001  & -0.0001
\end{bmatrix}, U_2 =\begin{bmatrix}
0  & 0\\
0 &  0
\end{bmatrix},\\
X_3 =& \begin{bmatrix}
1   &      0\\
0  &  1\\
0 &  -0.5\\
0  &  1\\
0   &      0\\
0   &      0
\end{bmatrix}, U_3=\begin{bmatrix}
0 &   0.0530\\
0  &  0.0531
\end{bmatrix}.\nonumber
\end{align}
For controller gain matrix $ K $ calculation, we take Fig.~\ref{algorithm1_perfect} (a)-(c)  for example and use Algorithm~\ref{alg:MaxiMinNSGA} with $ \bar \tau = 0.9 $ to obtain that $ K= \begin{bmatrix}
-0.0949  & -0.5737 &  -0.2657
\end{bmatrix} $ for agent $ 1 $ and $ 4 $, 
 $K= \begin{bmatrix}
-0.0953  & 0 &  -0.5383  &  0\\
0 &  -0.0953 &  0  & -0.5383
\end{bmatrix} $ for agent $ 2 $ and for agent $ 3 $:
\vspace{-0.5cm}
{\small  $$K= \begin{bmatrix}
	-0.0477 &  -0.0444 &   0.0171 &  -0.1478  & -0.1336 &   0.0096\\
	-0.0444 &  -0.0477 &  -0.0171  & -0.1336  & -0.1478 &  -0.0096
	\end{bmatrix}. $$ }
\vspace{-0.5cm}

Fig.~\ref{algorithm1_perfect} (a)-(c) demonstrate that {Assumption~\ref{assump_A0} is precise and} Algorithm~\ref{alg:MaxiMinNSGA} is available for MAS consensus tracking control as (i) errors are bounded with effects of external disturbances attenuated {during $ t\in [200,400] $; (ii) when $ t\in [0,200)$ without disturbances, $ \tilde{x}_{11}(t) $ is bounded (equals 0.2) in (a), zero in (b) and unbounded in (c) which verifies \eqref{chap_7_error3}   }. 

In the following, to better present the performance comparison, we do not add $ v_i(t) $.
Fig.~\ref{algorithm1_perfect} (d)-(f)
show that Algorithms~\ref{alg:MaxiMinNSGA2} can help systems endure larger input delay size.
{The value setting mechanism of $ \chi $ in the proposed  objective-function-transformed constraint~\eqref{constriant} is as follows. In Fig.~\ref{algorithm1_perfect} (e), we get $ \gamma/(2\delta \lambda_{\min}(P)) = 9.7842\exp+3 $ from Algorithms~\ref{alg:MaxiMinNSGA} for agent $ 2 $ which becomes unstable first. So we should choose $ \chi >9.7842\exp+3 $, e.g,  $ \chi =250000 $ in Fig.~\ref{algorithm1_perfect} (f). In fact, it is stable for $ \chi = 100000 $ with $ \bar \tau = 5.5 $, but not stable with $ \bar \tau = 5.6 $, then we choose $ \chi = 250000 $.}

\begin{figure}[!t]
	\centering
	\includegraphics[width=\hsize]{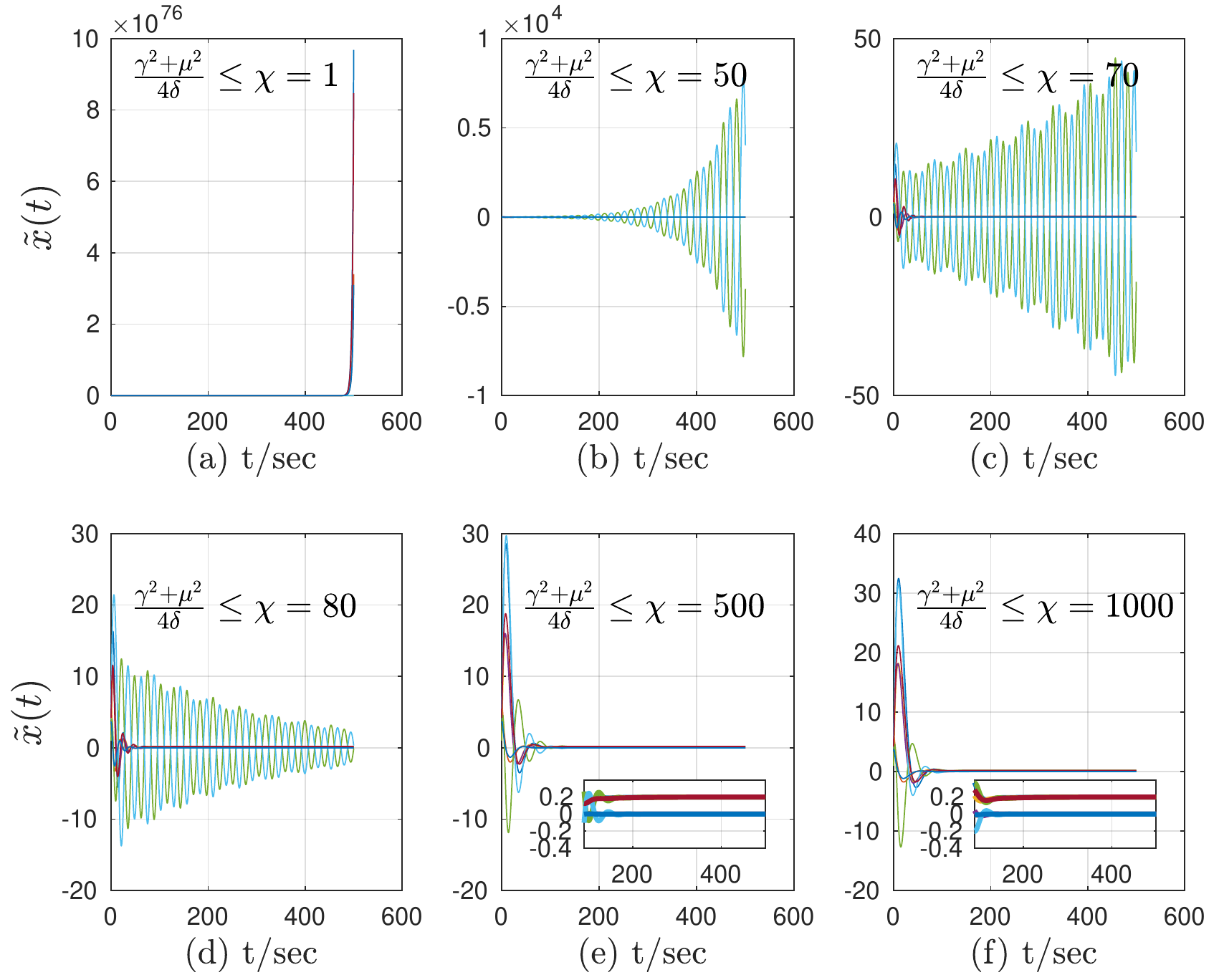}
	\caption{Influence of the value of $ \chi $ to MAS stability for Algorithms~\ref{alg:MaxiMinNSGA2} with a fixed delay size $ \bar \tau = 2.4 $.
	}
	\label{algorithm2_gammaComparison}
\end{figure}

Fig.~\ref{algorithm2_gammaComparison} gives more details about analyzing the influence of  constraint~\eqref{constriant}.
When $  \chi = 1 $ in Fig.~\ref{algorithm2_gammaComparison} (a), MAS is unstable,
which may be due to the  too strong  constraint for LMI variables $ \gamma $ and $ \mu $ in $ (\gamma^2 + \mu ^2) / (4\delta ) \le  1 $. Then, this constraint is relieved in Fig.~\ref{algorithm2_gammaComparison} (b)-(f)
where one can see  MAS becomes stable gradually.
Note that further alleviation of the constraint is not very helpful as there is nearly no performance difference between Fig.~\ref{algorithm2_gammaComparison} (e) and (f). This finding also verifies~\eqref{chap_7_error2}/\eqref{chap_7_error3}.
\begin{remark}
	Fig.~\ref{algorithm1_perfect} (e)-(f) and Fig.~\ref{algorithm2_gammaComparison} show when the closed-loop system is on the edge of stable/unstable state with delays, adjusting the constraint of $ (\gamma^2 + \mu ^2) / (4\delta ) \le \chi $ can improve the system ability of keeping stable, i.e., tuning $  \chi $ provides a freedom for a system to endure a larger delay size. {From Fig.~\ref{algorithm2_gammaComparison} and Fig.~\ref{algorithm1_perfect} (f), a basic rule is that for a larger delay size $ \bar \tau $, a larger $ \chi $ is required.}
	{For a fixed $ \chi $,} a fast way of finding a good bound for the delay is to use the bisection algorithm.
\end{remark}

\section{Conclusions and future directions}
\label{sec_conclusions}

This paper can address the heterogeneous and time-varying input and communication delays simultaneously by decoupling them when designing observers and using the Lyapunov-Krasovskii functional.
A new linear matrix inequality (LMI) is added to existing LMIs to construct an extended LMI approach to help an agent/a system endure larger delays compared to existing LMI solutions. 
Detailed analysis on how to obtain a larger delay upper bound and better robust control performance are also provided.
The proposed controllers and algorithms are without any integral term and thus, can be easily implemented in real applications. 
It is also easy to apply the proposed theory  to large-scale systems because (i) there is no requirement for global information, e.g., the eigenvalues of communication graph; (ii) the LMI dimension will not increase as agent number and delay number increase.

Future work will focus on designing observers that the state matrix can have eigenvalues with positive real parts and that  heterogeneous time-varying communication delays can be unknown.

\appendix
\section{Proof of Lemma~\ref{theorem_tracking_input}}\label{appendix_lemma1}
To facilitate the presentation of proof for $ \lim_{t \to \infty} \tilde \xi_i (t) =  0 $ exponentially, the following lemma is firstly needed.
\begin{lemma}[Theorem 2, \cite{moreau2004stability}]\label{lemma_morceau}
	Consider the linear system
	\begin{equation}\label{morceau_dynamics}
	\dot{x}(t) =  \underline{\text{diag}}(A(t))x(t) + (A(t) - \underline{\text{diag}}(A(t)) ) x(t-\tau)
	\end{equation}
	with $ x \in \mathbb{R}^{n}, \tau >0 $ and $  \underline{\text{diag}}(A(t)) $ is the obvious notation for the diagonal matrix obtained from $ A(t) $ by setting all off-diagonal entries equal to zero. Assume that the system matrix $ A(t) $ is a bounded and piecewise continuous function of time. Assume that, for every time $ t $, the system matrix is Metzler
	with zero row sums. If there is $ k \in \{1, \ldots, n\}, \delta >0 $ and $ T>0 $ such that for all $ t \in \mathbb{R} $, the $ \delta $-digraph\footnote{$ \delta $ is a threshold indicating that there exists an edge $ (i,j) $ in a digraph if and only if its corresponding weight $ a_{ij} >\delta $. } associated to $ \int_{t}^{t+T} A(s)ds $ has the property that all nodes may be reached from the node $ k $, then the equilibrium set of consensus states is uniformly exponentially stable. In particular, all components of any solution of~\eqref{morceau_dynamics} converge to a common value as $ t \rightarrow \infty $.
\end{lemma}

\begin{remark}
	\cite{7523319} extended Lemma~\ref{lemma_morceau} for the case of time-varying delays, i.e., $ \tau_{ij}(t) $ in [Lemma 3.1, \cite{7523319}].
	In this paper, Assumption~\ref{assumptiondirected} means the leader is the node $ k $ in Lemma~\ref{lemma_morceau}.
\end{remark}

Denote $ w _0 \coloneqq e^{-At} x_0 , w _i \coloneqq e^{-At} \xi_i, i \in \textbf{I}_1^N $. 
Then, from \eqref{chap_7_vl_dynamics}, \eqref{leader_observer} and based on $ e^{A\tau_{c_{ij}}}  x_0 (t- \tau_{c_{ij}} ) =  x_0 (t) $, we get
\begin{align}
\dot{w} _0(t) =& -A e^{-At} x_0 (t) + e^{-At} \dot x_0 (t) = 0, \\
\dot{w} _i(t) =& -A e^{-At}\xi_i (t)+ e^{-At}\dot{\xi}_i (t) \nonumber\\
=& -A e^{-At}\xi_i (t)+ e^{-At} \{ A\xi_i(t) \nonumber\\&+ \epsilon \sum_{j=1, j \neq i}^{N} a_{ij}[ e^{A\tau_{c_{ij}}} \xi_j (t- \tau_{c_{ij}} ) - \xi_i(t)] \nonumber\\
&+ \epsilon a_{i0} [ e^{A\tau_{c_{i0}}} x_0 (t- \tau_{c_{i0}} ) - \xi_i(t)  ] \}  \nonumber\\ 
=&  \epsilon \sum_{j=1, j \neq i}^{N} a_{ij}[e^{-A (t - \tau_{c_{ij}} )}  \xi_j (t- \tau_{c_{ij}} ) \nonumber\\& -\epsilon \sum_{j=0, j \neq i}^{N} a_{ij}e^{-A t}  \xi_i (t )  \nonumber\\&  +   \epsilon a_{i0}   e^{-A(t-\tau_{c_{i0}})} x_0 (t- \tau_{c_{i0}} ) \nonumber\\ 
=& -\epsilon \sum_{j \in \mathcal{ N}, j \neq i} l_{ij}  w_j (t- \tau_{c_{ij}} ) -\epsilon l_{ii}  w_i(t).
\end{align}
Denote $ \bar w_j(t)= [w_{0j}(t), w_{1j}(t), \ldots, w_{Nj}(t)]^T $ for each dimension $ j, j \in \textbf{I}_1^n $. Then, we have
\begin{align*}
\dot{\bar w} _j(t)  = \underline{\text{diag}}(-\epsilon\mathcal{L})\bar w_j(t) + (-\epsilon\mathcal{L} - \underline{\text{diag}}(-\epsilon\mathcal{L}) ) \bar w_j(t-\tau_c)
\end{align*}
where $ \bar w_j(t-\tau_c)   $ represents the corresponding delay term.
Based on the definition of Laplacian matrix $ \mathcal{L} $, from  Assumption~\ref{assumptiondirected}, one can see $ -\epsilon\mathcal{L} $ is Metzler with zero row sums. Thus, 
based on Lemma~\ref{lemma_morceau} with its extension for time-varying delay case in [Lemma 3.1, \cite{7523319}], we obtain $ w_{0j}(t)= w_{1j}(t)= \ldots= w_{Nj}(t), j \in \textbf{I}_1^n $ exponentially as $ t \rightarrow \infty $, which also means $  w_{0}(t)= w_{1}(t)= \ldots= w_{N}(t), t \rightarrow \infty $. Based on definitions of $ w_0(t) $ and $ w_i(t) $, we arrive at $ \lim_{t \to \infty} e^{-At} (\xi_i (t)-x_0 (t)) = \lim_{t \to \infty}  e^{-At} \tilde \xi_i (t) =  0 $. 

Denote $ \tilde{w}_i (t) \coloneqq e^{-At} \tilde \xi_i (t), \tilde w(t) = [ \tilde w_1^T(t), \ldots,  \tilde w^T_N(t) ]^T $.
The following follows the proof of Lemma 3.3 in~\cite{7523319}. Since $ \lim_{t \to \infty} \tilde{w} (t)  = 0 $ exponentially, there exist positive real numbers $ \alpha_1 $ and $ \beta $ such that 
\begin{align}
\| \tilde{w} (t)\| \le& \alpha_1 e^{-\beta t} \sup_{\theta \in [-\tau, 0]} \| \tilde{w}_t (\theta)\| \nonumber \\ 
=& \alpha_1 e^{-\beta t} \sup_{\theta \in [-\tau, 0]}  \| (I_N \otimes e^{-A \theta}) \tilde{\xi} (\theta)\|\nonumber \\ 
\le & \alpha e^{-\beta t},
\end{align}
where $ \alpha $ is some positive real number.
Based on Assumption~\ref{assump_A} that $ \mathrm{Re}(\lambda(A)) \le 0 $, there is a polynomial $ \Gamma (t) $ such that $ \| (I_N \otimes e^{A t}) \| \le \Gamma (t) $.
Then, 
\begin{equation}
\|  \tilde{\xi} (t)\| \le \| (I_N \otimes e^{A t}) \| \| \tilde{w} (t)\| \le \alpha e^{-\beta t} \Gamma (t).
\end{equation}
Therefore, $ \lim_{t \to \infty}  \tilde{\xi} (t) = 0$ exponentially. $ \blacksquare $


\section{Calculation of $ W $ in~\eqref{Lyapunov_W}}\label{appendixA}    
\vspace{-0.3cm}

Based on~\eqref{chap_7_x_tilde_Z1} and the LKF in Section~\ref{sec_mismatched}, we have
\begin{align}
W=&
2 \zeta^{T}P \dot{\zeta} + 2\delta \zeta^{T}P \zeta - \varpi^T\gamma\varpi
\nonumber\\&+ \zeta^{T}(S+Q) \zeta
- e^{-2\delta \bar{\tau}} \zeta^{T}(t-\bar{\tau})S\zeta(t-\bar{\tau}) \nonumber \\& - (1- \dot{\tau}(t)) e^{-2\delta \tau(t)} \zeta^{T}(t-\tau(t))Q\zeta(t-\tau(t))  
\nonumber\\& + \bar{\tau}^{2}  \dot{\zeta}^{T}(t)R \dot{\zeta}(t)  - \bar{\tau}   \int_{t-\bar{\tau}}^{t} e^{2\delta (s-t)} \dot{\zeta}^{T}(s)R\dot{\zeta}(s)\text{d}s \nonumber\\
\le & 2\zeta^{T}P \dot{\zeta} + 2\delta \zeta^{T}P \zeta \label{W_calculation} \\&- (1- \hat{\tau}) e^{-2\delta \bar{\tau}} \zeta^{T}(t-\tau(t))Q\zeta(t-\tau(t))
\nonumber\\&+ \zeta^{T}(S+Q) \zeta 
- e^{-2\delta \bar{\tau}} \zeta^{T}(t-\bar{\tau})S\zeta(t-\bar{\tau}) \nonumber \\
& + \bar{\tau}^{2}  \dot{\zeta}^{T}R \dot{\zeta}  - \bar{\tau}  e^{-2\delta \bar{\tau}} \int_{t-\bar{\tau}}^{t}  \dot{\zeta}^{T}(s)R\dot{\zeta}(s)\text{d}s - \varpi^T\gamma\varpi
\nonumber\\& +2 [\zeta^{T} P_2^T +  \dot{\zeta}^{T} P_3^T ] [A \zeta + BK \zeta(t-\tau(t)) +  \varpi - \dot{\zeta}].\nonumber
\end{align}
So the integral term $ \int_{t-\bar{\tau}}^{t}  \dot{\zeta}^{T}(s)R\dot{\zeta}(s)\text{d}s $  needs to be addressed. Applying Jensen's inequality, we have
\begin{align}
- &\bar{\tau} \int_{t-\bar{\tau}}^{t}  \dot{\zeta}^{T}(s)R\dot{\zeta}(s)ds \nonumber\\
=&     - \bar{\tau} \int_{t-\bar{\tau}}^{t-\tau (t)}  \dot{\zeta}^{T}(s)R\dot{\zeta}(s)ds  - \bar{\tau} \int_{t-\tau (t)}^{t}  \dot{\zeta}^{T}(s)R\dot{\zeta}(s)ds \nonumber\\
\le& \frac{- \bar{\tau}}{\bar{\tau}-\tau (t)} \int_{t-\bar{\tau}}^{t-\tau (t)}  \dot{\zeta}^{T}(s)ds \, R \int_{t-\bar{\tau}}^{t-\tau (t)} \dot{\zeta}(s)ds \nonumber\\&+ \frac{-\bar{\tau}}{\tau (t)} \int_{t-\tau (t)}^{t}  \dot{\zeta}^{T}(s)ds \, R \int_{t-\tau (t)}^{t} \dot{\zeta}(s)ds \nonumber\\
=& \frac{- \bar{\tau}}{\bar{\tau}-\tau (t)} \vartheta_{2}^{T}R\vartheta_{2} + \frac{-\bar{\tau}}{\tau (t)} \vartheta_{1}^{T}R\vartheta_{1} \nonumber\\
=& - \begin{bmatrix}
\vartheta_{1}^T &  \vartheta_{2}^T
\end{bmatrix} \begin{bmatrix}
\frac{1}{\frac{\tau (t)}{\bar{\tau}}}R & 0\\
0& \frac{1}{\frac{\bar{\tau}-\tau (t)}{\bar{\tau}}}R 
\end{bmatrix}\begin{bmatrix}
\vartheta_{1} \\  \vartheta_{2}
\end{bmatrix}, \label{chap_6_lyakro2}  
\end{align}
where $ \vartheta_{1} =\zeta(t)- \zeta(t-\tau (t)) $ and $ \vartheta_{2} = \zeta(t-\tau (t)) - \zeta(t-\bar{\tau}) $. Here, when $ \tau (t) \rightarrow 0 $, there exists the following limit:
\begin{equation*}
\begin{aligned}
\lim_{\tau (t) \to 0} \frac{-\bar{\tau}}{\tau (t)} \vartheta_{1}^{T}R\vartheta_{1} =& -\bar{\tau} \lim_{\tau (t) \to 0} \tau (t) \frac{\zeta^T(t)- \zeta^T(t-\tau (t))}{\tau (t)} \\& \times R \frac{\zeta(t)- \zeta(t-\tau (t))}{\tau (t)}\\
=& -\bar{\tau} \lim_{\tau (t) \to 0} \tau (t) \dot{\zeta}^T(t) R \dot{\zeta}(t)\\=&0.
\end{aligned}
\end{equation*}
Similarly, when $ \tau (t) \rightarrow \bar{\tau} $, there exists the following limit:
\begin{equation*}
\begin{aligned}
\lim_{\tau (t) \to \bar{\tau}} \frac{- \bar{\tau}}{\bar{\tau}-\tau (t)} \vartheta_{2}^{T}R\vartheta_{2} =0.
\end{aligned}
\end{equation*}
Now we are ready to borrow the following lemma.
\begin{lemma}[\cite{fridman2014introduction}]\label{chap_6_lemma_fridman}
	Let $ R_{1} \in \mathbb{R}^{n_{1} \times n_{1}}, \ldots,\\ R_{N} \in \mathbb{R}^{n_{N} \times n_{N}} $ be positive matrices, Then for all $ \vartheta_{1}\in \mathbb{R}^{n_{1}}, \ldots, \vartheta_{N}\in \mathbb{R}^{n_{N}}  $, for all $ \alpha_{i} >0 $ with $ \displaystyle\sum_{i} \alpha_{i} =1$ and for all $ S_{ij} \in \mathbb{R}^{n_{i} \times n_{j}}, i \in \textbf{I}_1^N, j \in \textbf{I}_1^{i-1}  $ such that 
	\begin{equation}
	\begin{bmatrix}
	R_{i} & S_{ij}\\ * & R_{j}
	\end{bmatrix}\ge 0,
	\end{equation}
	the following inequality holds:
	\begin{equation}
	\sum_{i=1}^{N}\frac{1}{\alpha_{i}} \vartheta_{i}^{T}R_{i}\vartheta_{i} \ge \begin{bmatrix}
	\vartheta_{1}\\ \vartheta_{2}\\ \vdots\\ \vartheta_{N}
	\end{bmatrix}^{T}
	\begin{bmatrix}
	R_1 & S_{12} & \cdots & S_{1N} \\
	* & R_2 & \cdots & S_{2N} \\
	*& * & \ddots & \vdots \\
	* & * & \cdots & R_N    
	\end{bmatrix}
	\begin{bmatrix}
	\vartheta_{1}\\ \vartheta_{2}\\ \vdots\\ \vartheta_{N}
	\end{bmatrix}.
	\end{equation}
\end{lemma}
By applying Lemma~\ref{chap_6_lemma_fridman}, \eqref{chap_6_lyakro2} becomes
\begin{equation}\label{chap_6_lyakro3}
\begin{aligned}
- \bar{\tau} \int_{t-\bar{\tau}}^{t}  \dot{\zeta}^{T}(s)R\dot{\zeta}(s)ds
\le - \begin{bmatrix}
\vartheta_{1}^T &  \vartheta_{2}^T
\end{bmatrix} \begin{bmatrix}
R & S_{12}\\ * & R
\end{bmatrix}\begin{bmatrix}
\vartheta_{1} \\  \vartheta_{2}
\end{bmatrix}.
\end{aligned}
\end{equation}

Integrating \eqref{chap_6_lyakro3} into \eqref{W_calculation}, we have
\begin{equation}\label{chap_7_lyakro8}
W \le \bar{\zeta}^T\Phi_1\bar{\zeta}
\end{equation}
where $ \bar{\zeta}(t) =  \text{col}(\zeta(t), \dot{\zeta}(t), \zeta(t-\bar{\tau}), \zeta(t-\tau(t)), \varpi(t))$, 
{\small
\begin{align}\label{lmi_phi4}
\Phi_1 &=\begin{bmatrix}
\Phi_1 (1,1)&\Phi_1 (1,2) & e^{-2\delta \bar{\tau}} S_{12} & \Phi_1 (1,4) &  P_2^T \\
*&\Phi_1 (2,2)&0&  P_3^T BK& P_3^T \\
*&*&\Phi_1 (3,3)& \Phi_1 (3,4)& 0\\
*&*&*&\Phi_1 (4,4) &0\\
*&*&*&*&-\gamma I_n
\end{bmatrix},\\
\Phi_1& (1,1) =2\delta P+S+Q-e^{-2\delta \bar{\tau}} R+P_2^T A+A^T  P_2, \nonumber\\
\Phi_1& (1,2) = P-P_2^T+A^T  P_3, \Phi_1 (3,4)=e^{-2\delta \bar{\tau}} (R-S_{12}^T),\nonumber\\
\Phi_1& (1,4) = P_2^T BK+ e^{-2\delta \bar{\tau}} (R-S_{12}),\nonumber\\
\Phi_1& (2,2) = \bar{\tau}^{2}R-P_3 - P_3^T,
\Phi_1 (3,3) = -e^{-2\delta \bar{\tau}}(S+R),\nonumber\\
\Phi_1& (4,4) = -(1-\hat{\tau})e^{-2\delta \bar{\tau}}Q+ e^{-2\delta \bar{\tau}} (-2R+S_{12}+S_{12}^T),\nonumber
\end{align}
}\noindent
where $\{P_2, P_3, S_{12}\} \in  \mathbb{R}^{n\times n} $ will be decided later.
The inequality~\eqref{chap_7_lyakro8} comes from $ \dot{\tau}(t) \le \hat{\tau} $ in Assumption~\ref{assump_delay}, the Jensen's inequality and Lemma 3.4 in~\cite{fridman2014introduction_book} where the matrix $ S_{12} $ is introduced to satisfy {\small
\begin{equation}\label{S12}
\Phi_1^{'}  \succeq 0.
\end{equation}}\noindent
where $ \Phi_1^{'} =  $ {\tiny $ \begin{bmatrix}
	R & S_{12}\\ * & R
	\end{bmatrix} $}.
For more mathematical  details about calculating \eqref{chap_7_lyakro8} and \eqref{lmi_phi4}, please refer to Eqs.~(6.5)-(6.20) in~\cite{jiang2018fully}.
The last term (called the descriptor method in~\cite{fridman2014introduction_book} where $ P_2, P_3 $ are introduced) in inequality~\eqref{chap_7_lyakro8} is identically zero, which comes directly from~\eqref{chap_7_x_tilde_Z1}. \hfill $ \blacksquare $


\bibliographystyle{apacite} 
\bibliography{mybibfile}           

\begin{thebibliography}{}

\bibitem [\protect \citeauthoryear {%
Ahmed%
, Khan%
, Saeed%
\BCBL {}\ \BBA {} Zhang%
}{%
Ahmed%
\ \protect \BOthers {.}}{%
{\protect \APACyear {2020}}%
}]{%
ahmed2020consensus}
\APACinsertmetastar {%
ahmed2020consensus}%
\begin{APACrefauthors}%
Ahmed, Z.%
, Khan, M\BPBI M.%
, Saeed, M\BPBI A.%
\BCBL {}\ \BBA {} Zhang, W.%
\end{APACrefauthors}%
\unskip\
\newblock
\APACrefYearMonthDay{2020}{}{}.
\newblock
{\BBOQ}\APACrefatitle {Consensus control of multi-agent systems with input and
  communication delay: A frequency domain perspective} {Consensus control of
  multi-agent systems with input and communication delay: A frequency domain
  perspective}.{\BBCQ}
\newblock
\APACjournalVolNumPages{ISA Transactions}{101}{}{69--77}.
\PrintBackRefs{\CurrentBib}

\bibitem [\protect \citeauthoryear {%
{Besan{\c c}on}%
, {Georges}%
\BCBL {}\ \BBA {} {Benayache}%
}{%
{Besan{\c c}on}%
\ \protect \BOthers {.}}{%
{\protect \APACyear {2007}}%
}]{%
Besancon2007}
\APACinsertmetastar {%
Besancon2007}%
\begin{APACrefauthors}%
{Besan{\c c}on}, G.%
, {Georges}, D.%
\BCBL {}\ \BBA {} {Benayache}, Z.%
\end{APACrefauthors}%
\unskip\
\newblock
\APACrefYearMonthDay{2007}{}{}.
\newblock
{\BBOQ}\APACrefatitle {Asymptotic state prediction for continuous-time systems
  with delayed input and application to control} {Asymptotic state prediction
  for continuous-time systems with delayed input and application to
  control}.{\BBCQ}
\newblock
\BIn{} \APACrefbtitle {2007 {E}uropean {C}ontrol {C}onference} {2007 {E}uropean
  {C}ontrol {C}onference}\ (\BPGS\ 1786--1791).
\PrintBackRefs{\CurrentBib}

\bibitem [\protect \citeauthoryear {%
Cai%
, Lewis%
, Hu%
\BCBL {}\ \BBA {} Huang%
}{%
Cai%
\ \protect \BOthers {.}}{%
{\protect \APACyear {2017}}%
}]{%
cai2017adaptive}
\APACinsertmetastar {%
cai2017adaptive}%
\begin{APACrefauthors}%
Cai, H.%
, Lewis, F\BPBI L.%
, Hu, G.%
\BCBL {}\ \BBA {} Huang, J.%
\end{APACrefauthors}%
\unskip\
\newblock
\APACrefYearMonthDay{2017}{}{}.
\newblock
{\BBOQ}\APACrefatitle {The adaptive distributed observer approach to the
  cooperative output regulation of linear multi-agent systems} {The adaptive
  distributed observer approach to the cooperative output regulation of linear
  multi-agent systems}.{\BBCQ}
\newblock
\APACjournalVolNumPages{Automatica}{75}{}{299--305}.
\PrintBackRefs{\CurrentBib}

\bibitem [\protect \citeauthoryear {%
De%
, Sahoo%
\BCBL {}\ \BBA {} Wahi%
}{%
De%
\ \protect \BOthers {.}}{%
{\protect \APACyear {2018}}%
}]{%
de2018trajectory}
\APACinsertmetastar {%
de2018trajectory}%
\begin{APACrefauthors}%
De, S.%
, Sahoo, S\BPBI R.%
\BCBL {}\ \BBA {} Wahi, P.%
\end{APACrefauthors}%
\unskip\
\newblock
\APACrefYearMonthDay{2018}{}{}.
\newblock
{\BBOQ}\APACrefatitle {Trajectory tracking control with heterogeneous input
  delay in multi-agent system} {Trajectory tracking control with heterogeneous
  input delay in multi-agent system}.{\BBCQ}
\newblock
\APACjournalVolNumPages{Journal of Intelligent \& Robotic
  Systems}{92}{3-4}{521--544}.
\PrintBackRefs{\CurrentBib}

\bibitem [\protect \citeauthoryear {%
Fridman%
}{%
Fridman%
}{%
{\protect \APACyear {2014}}%
{\protect \APACexlab {{\protect \BCnt {1}}}}}]{%
fridman2014introduction}
\APACinsertmetastar {%
fridman2014introduction}%
\begin{APACrefauthors}%
Fridman, E.%
\end{APACrefauthors}%
\unskip\
\newblock
\APACrefYearMonthDay{2014{\protect \BCnt {1}}}{}{}.
\newblock
{\BBOQ}\APACrefatitle {Introduction to time-delay and sampled-data systems}
  {Introduction to time-delay and sampled-data systems}.{\BBCQ}
\newblock
\BIn{} \APACrefbtitle {2014 {E}uropean {C}ontrol {C}onference} {2014 {E}uropean
  {C}ontrol {C}onference}\ (\BPGS\ 1428--1433).
\PrintBackRefs{\CurrentBib}

\bibitem [\protect \citeauthoryear {%
Fridman%
}{%
Fridman%
}{%
{\protect \APACyear {2014}}%
{\protect \APACexlab {{\protect \BCnt {2}}}}}]{%
fridman2014introduction_book}
\APACinsertmetastar {%
fridman2014introduction_book}%
\begin{APACrefauthors}%
Fridman, E.%
\end{APACrefauthors}%
\unskip\
\newblock
\APACrefYear{2014{\protect \BCnt {2}}}.
\newblock
\APACrefbtitle {Introduction to Time-Delay Systems: Analysis and Control}
  {Introduction to time-delay systems: Analysis and control}.
\newblock
\APACaddressPublisher{}{Springer}.
\PrintBackRefs{\CurrentBib}

\bibitem [\protect \citeauthoryear {%
Fridman%
\ \BBA {} Dambrine%
}{%
Fridman%
\ \BBA {} Dambrine%
}{%
{\protect \APACyear {2009}}%
}]{%
fridman2009control}
\APACinsertmetastar {%
fridman2009control}%
\begin{APACrefauthors}%
Fridman, E.%
\BCBT {}\ \BBA {} Dambrine, M.%
\end{APACrefauthors}%
\unskip\
\newblock
\APACrefYearMonthDay{2009}{}{}.
\newblock
{\BBOQ}\APACrefatitle {Control under quantization, saturation and delay: An
  {LMI} approach} {Control under quantization, saturation and delay: An {LMI}
  approach}.{\BBCQ}
\newblock
\APACjournalVolNumPages{Automatica}{45}{10}{2258--2264}.
\PrintBackRefs{\CurrentBib}

\bibitem [\protect \citeauthoryear {%
Hespanha%
}{%
Hespanha%
}{%
{\protect \APACyear {2018}}%
}]{%
hespanha2018linear}
\APACinsertmetastar {%
hespanha2018linear}%
\begin{APACrefauthors}%
Hespanha, J\BPBI P.%
\end{APACrefauthors}%
\unskip\
\newblock
\APACrefYear{2018}.
\newblock
\APACrefbtitle {Linear Systems Theory} {Linear systems theory}.
\newblock
\APACaddressPublisher{}{Princeton university press}.
\PrintBackRefs{\CurrentBib}

\bibitem [\protect \citeauthoryear {%
Hou%
, Fu%
, Zhang%
\BCBL {}\ \BBA {} Wu%
}{%
Hou%
\ \protect \BOthers {.}}{%
{\protect \APACyear {2017}}%
}]{%
hou2017consensus}
\APACinsertmetastar {%
hou2017consensus}%
\begin{APACrefauthors}%
Hou, W.%
, Fu, M.%
, Zhang, H.%
\BCBL {}\ \BBA {} Wu, Z.%
\end{APACrefauthors}%
\unskip\
\newblock
\APACrefYearMonthDay{2017}{}{}.
\newblock
{\BBOQ}\APACrefatitle {Consensus conditions for general second-order
  multi-agent systems with communication delay} {Consensus conditions for
  general second-order multi-agent systems with communication delay}.{\BBCQ}
\newblock
\APACjournalVolNumPages{Automatica}{75}{}{293--298}.
\PrintBackRefs{\CurrentBib}

\bibitem [\protect \citeauthoryear {%
Huang%
}{%
Huang%
}{%
{\protect \APACyear {2004}}%
}]{%
huang2004nonlinear}
\APACinsertmetastar {%
huang2004nonlinear}%
\begin{APACrefauthors}%
Huang, J.%
\end{APACrefauthors}%
\unskip\
\newblock
\APACrefYear{2004}.
\newblock
\APACrefbtitle {Nonlinear Output Regulation: Theory and Applications}
  {Nonlinear output regulation: Theory and applications}.
\newblock
\APACaddressPublisher{}{SIAM}.
\PrintBackRefs{\CurrentBib}

\bibitem [\protect \citeauthoryear {%
Jiang%
}{%
Jiang%
}{%
{\protect \APACyear {2018}}%
}]{%
jiang2018fully}
\APACinsertmetastar {%
jiang2018fully}%
\begin{APACrefauthors}%
Jiang, W.%
\end{APACrefauthors}%
\unskip\
\newblock
\APACrefYear{2018}.
\newblock
\APACrefbtitle {Fully Distributed Time-varying Formation and Containment
  Control for Multi-agent / Multi-robot Systems} {Fully distributed
  time-varying formation and containment control for multi-agent / multi-robot
  systems}.
\newblock
Ph.D. thesis, Automatic Control Engineering. Ecole Centrale de Lille. English.
  tel-02097750.
\PrintBackRefs{\CurrentBib}

\bibitem [\protect \citeauthoryear {%
{Jiang}%
, {Chen}%
\BCBL {}\ \BBA {} {Charalambous}%
}{%
{Jiang}%
\ \protect \BOthers {.}}{%
{\protect \APACyear {2021}}%
}]{%
jiang_LCCS}
\APACinsertmetastar {%
jiang_LCCS}%
\begin{APACrefauthors}%
{Jiang}, W.%
, {Chen}, Y.%
\BCBL {}\ \BBA {} {Charalambous}, T.%
\end{APACrefauthors}%
\unskip\
\newblock
\APACrefYearMonthDay{2021}{}{}.
\newblock
{\BBOQ}\APACrefatitle {Consensus of General Linear Multi-Agent Systems With
  Heterogeneous Input and Communication Delays} {Consensus of general linear
  multi-agent systems with heterogeneous input and communication
  delays}.{\BBCQ}
\newblock
\APACjournalVolNumPages{IEEE Control Systems Letters}{5}{3}{851-856}.
\PrintBackRefs{\CurrentBib}

\bibitem [\protect \citeauthoryear {%
{Jiang}%
, {Wen}%
, {Peng}%
, {Huang}%
\BCBL {}\ \BBA {} {Rahmani}%
}{%
{Jiang}%
\ \protect \BOthers {.}}{%
{\protect \APACyear {2019}}%
}]{%
jiang_tac}
\APACinsertmetastar {%
jiang_tac}%
\begin{APACrefauthors}%
{Jiang}, W.%
, {Wen}, G.%
, {Peng}, Z.%
, {Huang}, T.%
\BCBL {}\ \BBA {} {Rahmani}, A.%
\end{APACrefauthors}%
\unskip\
\newblock
\APACrefYearMonthDay{2019}{}{}.
\newblock
{\BBOQ}\APACrefatitle {Fully Distributed Formation-Containment Control of
  Heterogeneous Linear Multiagent Systems} {Fully distributed
  formation-containment control of heterogeneous linear multiagent
  systems}.{\BBCQ}
\newblock
\APACjournalVolNumPages{IEEE Transactions on Automatic
  Control}{64}{9}{3889-3896}.
\PrintBackRefs{\CurrentBib}

\bibitem [\protect \citeauthoryear {%
L{\'e}chapp{\'e}%
, {Moulay}%
\BCBL {}\ \BBA {} {Plestan}%
}{%
L{\'e}chapp{\'e}%
\ \protect \BOthers {.}}{%
{\protect \APACyear {2016}}%
}]{%
Lechappe2016}
\APACinsertmetastar {%
Lechappe2016}%
\begin{APACrefauthors}%
L{\'e}chapp{\'e}, V.%
, {Moulay}, E.%
\BCBL {}\ \BBA {} {Plestan}, F.%
\end{APACrefauthors}%
\unskip\
\newblock
\APACrefYearMonthDay{2016}{}{}.
\newblock
{\BBOQ}\APACrefatitle {Dynamic observation-prediction for LTI systems with a
  time-varying delay in the input} {Dynamic observation-prediction for lti
  systems with a time-varying delay in the input}.{\BBCQ}
\newblock
\BIn{} \APACrefbtitle {55th {IEEE} {C}onference on {D}ecision and {C}ontrol}
  {55th {IEEE} {C}onference on {D}ecision and {C}ontrol}\ (\BPGS\ 2302--2307).
\PrintBackRefs{\CurrentBib}

\bibitem [\protect \citeauthoryear {%
Liu%
, Fridman%
\BCBL {}\ \BBA {} Xia%
}{%
Liu%
\ \protect \BOthers {.}}{%
{\protect \APACyear {2020}}%
}]{%
liu2020networked}
\APACinsertmetastar {%
liu2020networked}%
\begin{APACrefauthors}%
Liu, K.%
, Fridman, E.%
\BCBL {}\ \BBA {} Xia, Y.%
\end{APACrefauthors}%
\unskip\
\newblock
\APACrefYear{2020}.
\newblock
\APACrefbtitle {Networked Control Under Communication Constraints: A Time-Delay
  Approach} {Networked control under communication constraints: A time-delay
  approach}.
\newblock
\APACaddressPublisher{}{Springer}.
\PrintBackRefs{\CurrentBib}

\bibitem [\protect \citeauthoryear {%
{Lu}%
\ \BBA {} {Liu}%
}{%
{Lu}%
\ \BBA {} {Liu}%
}{%
{\protect \APACyear {2017}}%
}]{%
7523319}
\APACinsertmetastar {%
7523319}%
\begin{APACrefauthors}%
{Lu}, M.%
\BCBT {}\ \BBA {} {Liu}, L.%
\end{APACrefauthors}%
\unskip\
\newblock
\APACrefYearMonthDay{2017}{}{}.
\newblock
{\BBOQ}\APACrefatitle {Distributed Feedforward Approach to Cooperative Output
  Regulation Subject to Communication Delays and Switching Networks}
  {Distributed feedforward approach to cooperative output regulation subject to
  communication delays and switching networks}.{\BBCQ}
\newblock
\APACjournalVolNumPages{IEEE Transactions on Automatic
  Control}{62}{4}{1999-2005}.
\PrintBackRefs{\CurrentBib}

\bibitem [\protect \citeauthoryear {%
Moreau%
}{%
Moreau%
}{%
{\protect \APACyear {2004}}%
}]{%
moreau2004stability}
\APACinsertmetastar {%
moreau2004stability}%
\begin{APACrefauthors}%
Moreau, L.%
\end{APACrefauthors}%
\unskip\
\newblock
\APACrefYearMonthDay{2004}{}{}.
\newblock
{\BBOQ}\APACrefatitle {Stability of continuous-time distributed consensus
  algorithms} {Stability of continuous-time distributed consensus
  algorithms}.{\BBCQ}
\newblock
\BIn{} \APACrefbtitle {43rd {IEEE} {C}onference on {D}ecision and {C}ontrol}
  {43rd {IEEE} {C}onference on {D}ecision and {C}ontrol}\ (\BPGS\ 3998--4003).
\PrintBackRefs{\CurrentBib}

\bibitem [\protect \citeauthoryear {%
M{\"u}nz%
, Papachristodoulou%
\BCBL {}\ \BBA {} Allg{\"o}wer%
}{%
M{\"u}nz%
\ \protect \BOthers {.}}{%
{\protect \APACyear {2010}}%
}]{%
munz2010delay}
\APACinsertmetastar {%
munz2010delay}%
\begin{APACrefauthors}%
M{\"u}nz, U.%
, Papachristodoulou, A.%
\BCBL {}\ \BBA {} Allg{\"o}wer, F.%
\end{APACrefauthors}%
\unskip\
\newblock
\APACrefYearMonthDay{2010}{}{}.
\newblock
{\BBOQ}\APACrefatitle {Delay robustness in consensus problems} {Delay
  robustness in consensus problems}.{\BBCQ}
\newblock
\APACjournalVolNumPages{Automatica}{46}{8}{1252--1265}.
\PrintBackRefs{\CurrentBib}

\bibitem [\protect \citeauthoryear {%
Najafi%
, Hosseinnia%
, Sheikholeslam%
\BCBL {}\ \BBA {} Karimadini%
}{%
Najafi%
\ \protect \BOthers {.}}{%
{\protect \APACyear {2013}}%
}]{%
najafi2013closed}
\APACinsertmetastar {%
najafi2013closed}%
\begin{APACrefauthors}%
Najafi, M.%
, Hosseinnia, S.%
, Sheikholeslam, F.%
\BCBL {}\ \BBA {} Karimadini, M.%
\end{APACrefauthors}%
\unskip\
\newblock
\APACrefYearMonthDay{2013}{}{}.
\newblock
{\BBOQ}\APACrefatitle {Closed-loop control of dead time systems via sequential
  sub-predictors} {Closed-loop control of dead time systems via sequential
  sub-predictors}.{\BBCQ}
\newblock
\APACjournalVolNumPages{International Journal of Control}{86}{4}{599--609}.
\PrintBackRefs{\CurrentBib}

\bibitem [\protect \citeauthoryear {%
Olfati-Saber%
\ \BBA {} Murray%
}{%
Olfati-Saber%
\ \BBA {} Murray%
}{%
{\protect \APACyear {2004}}%
}]{%
olfati-saber_consensus_2004}
\APACinsertmetastar {%
olfati-saber_consensus_2004}%
\begin{APACrefauthors}%
Olfati-Saber, R.%
\BCBT {}\ \BBA {} Murray, R\BPBI M.%
\end{APACrefauthors}%
\unskip\
\newblock
\APACrefYearMonthDay{2004}{}{}.
\newblock
{\BBOQ}\APACrefatitle {Consensus problems in networks of agents with switching
  topology and time-delays} {Consensus problems in networks of agents with
  switching topology and time-delays}.{\BBCQ}
\newblock
\APACjournalVolNumPages{IEEE Transactions on Automatic
  Control}{49}{9}{118--173}.
\PrintBackRefs{\CurrentBib}

\bibitem [\protect \citeauthoryear {%
{Sun}%
\ \BBA {} {Wang}%
}{%
{Sun}%
\ \BBA {} {Wang}%
}{%
{\protect \APACyear {2009}}%
}]{%
5109512}
\APACinsertmetastar {%
5109512}%
\begin{APACrefauthors}%
{Sun}, Y\BPBI G.%
\BCBT {}\ \BBA {} {Wang}, L.%
\end{APACrefauthors}%
\unskip\
\newblock
\APACrefYearMonthDay{2009}{}{}.
\newblock
{\BBOQ}\APACrefatitle {Consensus of Multi-Agent Systems in Directed Networks
  With Nonuniform Time-Varying Delays} {Consensus of multi-agent systems in
  directed networks with nonuniform time-varying delays}.{\BBCQ}
\newblock
\APACjournalVolNumPages{IEEE Transactions on Automatic
  Control}{54}{7}{1607-1613}.
\PrintBackRefs{\CurrentBib}

\bibitem [\protect \citeauthoryear {%
{Tian}%
\ \BBA {} {Liu}%
}{%
{Tian}%
\ \BBA {} {Liu}%
}{%
{\protect \APACyear {2008}}%
}]{%
4639466}
\APACinsertmetastar {%
4639466}%
\begin{APACrefauthors}%
{Tian}, Y.%
\BCBT {}\ \BBA {} {Liu}, C.%
\end{APACrefauthors}%
\unskip\
\newblock
\APACrefYearMonthDay{2008}{}{}.
\newblock
{\BBOQ}\APACrefatitle {Consensus of Multi-Agent Systems With Diverse Input and
  Communication Delays} {Consensus of multi-agent systems with diverse input
  and communication delays}.{\BBCQ}
\newblock
\APACjournalVolNumPages{IEEE Transactions on Automatic
  Control}{53}{9}{2122-2128}.
\PrintBackRefs{\CurrentBib}

\bibitem [\protect \citeauthoryear {%
{Van Assche}%
, {Dambrine}%
, {Lafay}%
\BCBL {}\ \BBA {} {Richard}%
}{%
{Van Assche}%
\ \protect \BOthers {.}}{%
{\protect \APACyear {1999}}%
}]{%
833279}
\APACinsertmetastar {%
833279}%
\begin{APACrefauthors}%
{Van Assche}, V.%
, {Dambrine}, M.%
, {Lafay}, J\BPBI F.%
\BCBL {}\ \BBA {} {Richard}, J\BPBI P.%
\end{APACrefauthors}%
\unskip\
\newblock
\APACrefYearMonthDay{1999}{}{}.
\newblock
{\BBOQ}\APACrefatitle {Some problems arising in the implementation of
  distributed-delay control laws} {Some problems arising in the implementation
  of distributed-delay control laws}.{\BBCQ}
\newblock
\BIn{} \APACrefbtitle {38th {IEEE} {C}onference on {D}ecision and {C}ontrol}
  {38th {IEEE} {C}onference on {D}ecision and {C}ontrol}\ (\BPGS\ 4668--4672).
\PrintBackRefs{\CurrentBib}

\bibitem [\protect \citeauthoryear {%
{Xu}%
, {Liu}%
\BCBL {}\ \BBA {} {Feng}%
}{%
{Xu}%
\ \protect \BOthers {.}}{%
{\protect \APACyear {2018}}%
}]{%
xu2017consensus}
\APACinsertmetastar {%
xu2017consensus}%
\begin{APACrefauthors}%
{Xu}, X.%
, {Liu}, L.%
\BCBL {}\ \BBA {} {Feng}, G.%
\end{APACrefauthors}%
\unskip\
\newblock
\APACrefYearMonthDay{2018}{}{}.
\newblock
{\BBOQ}\APACrefatitle {Consensus of Discrete-Time Linear Multiagent Systems
  With Communication, Input and Output Delays} {Consensus of discrete-time
  linear multiagent systems with communication, input and output
  delays}.{\BBCQ}
\newblock
\APACjournalVolNumPages{IEEE Transactions on Automatic
  Control}{63}{2}{492-497}.
\PrintBackRefs{\CurrentBib}

\bibitem [\protect \citeauthoryear {%
Zhou%
\ \BBA {} Lin%
}{%
Zhou%
\ \BBA {} Lin%
}{%
{\protect \APACyear {2014}}%
}]{%
zhou2014consensus}
\APACinsertmetastar {%
zhou2014consensus}%
\begin{APACrefauthors}%
Zhou, B.%
\BCBT {}\ \BBA {} Lin, Z.%
\end{APACrefauthors}%
\unskip\
\newblock
\APACrefYearMonthDay{2014}{}{}.
\newblock
{\BBOQ}\APACrefatitle {Consensus of high-order multi-agent systems with large
  input and communication delays} {Consensus of high-order multi-agent systems
  with large input and communication delays}.{\BBCQ}
\newblock
\APACjournalVolNumPages{Automatica}{50}{2}{452--464}.
\PrintBackRefs{\CurrentBib}

\end{thebibliography}

\end{document}